\documentclass[letterpaper,twocolumn,10pt]{article}
\usepackage{usenix2019_v3}

\usepackage{amsmath}
\usepackage{capt-of}
\usepackage{color, colortbl}
\usepackage{enumitem}
\usepackage{multirow}
\usepackage{pifont}
\usepackage{tikz}

\graphicspath{{figures/}}

\usepackage{titlesec}
\titlespacing\section{0pt}{3pt plus 0pt minus 1pt}{3pt plus 0pt minus 1pt}
\titlespacing\subsection{0pt}{3pt plus 0pt minus 1pt}{3pt plus 0pt minus 1pt}
\titlespacing\subsubsection{0pt}{1pt plus 2pt minus 0pt}{1pt plus 2pt minus 2pt}

\newcommand{\needref}[1]{\textbf{\color{red}[Need Ref]}}

\newcommand{\dparfirst}{Targeted Tamper-Evident Routing (T-TER)}
\newcommand{\dpar}{T-TER}

\definecolor{GreenHLight}{RGB}{21,176,79}
\definecolor{RedHLight}{RGB}{250,34,0}
\definecolor{BlueHLight}{RGB}{53,176,239}
\definecolor{YellowHLight}{RGB}{251,217,102}

\newcommand{\figline}{\vspace{3pt}\hrulefill}
\newcommand{\cmark}{\ding{51}}%
\newcommand{\xmark}{\ding{55}}%

\begin{document}

\title{T-TER: Defeating A2 Trojans with Targeted Tamper-Evident Routing}

\author{
{\rm Timothy Trippel\thanks{Work completed at MIT Licoln Laboratory.} , Kang G. Shin}\\
Computer Science \& Engineering \\
University of Michigan \\
\{trippel,kgshin\}@umich.edu
\and
{\rm Kevin B. Bush}\\
Cyber Physical Systems \\
MIT Lincoln Laboratory \\
kevin.bush@ll.mit.edu
\and
{\rm Matthew Hicks\color{green!80!black}\footnotemark[1]
\thanks{Corresponding faculty author}}\\
Computer Science \\
Virginia Tech \\
mdhicks2@vt.edu
} 

\maketitle

\begin{abstract}

Since the inception of the Integrated Circuit (IC), the size of the transistors used to construct them has continually shrunk. While this advancement significantly improves computing capability, fabrication costs have skyrocketed. As a result, most IC designers must now outsource fabrication. Outsourcing, however, presents a security threat: comprehensive post-fabrication inspection is infeasible given the size of modern ICs, so it is nearly impossible to know if the foundry has altered the original design during fabrication (i.e., inserted a hardware Trojan). Defending against a foundry-side adversary is challenging because---even with as few as two gates---hardware Trojans can completely undermine 
software security. Researchers have attempted to both \textit{detect} and \textit{prevent} foundry-side attacks, but all existing defenses are ineffective against Trojans 
with footprints of a few gates or less.

We present \dparfirst{}, a \emph{preventive} layout-level defense against untrusted foundries, capable of thwarting the insertion of even the stealthiest hardware Trojans. \dpar{} is \textit{directed} and \textit{routing-centric}: it prevents foundry-side attackers from routing Trojan wires to, or directly adjacent to, security-critical wires by shielding them with guard wires. Unlike shield wires commonly deployed for cross-talk reduction, \dpar{} guard wires pose an additional technical challenge: they must be tamper-evident in both the digital (deletion attacks) and analog (move and jog attacks) domains. We address this challenge by developing a class of {\em designed-in} guard wires, that are added to the design specifically to protect security-critical wires. \dpar{}'s guard wires incur minimal overhead, scale with design complexity, and provide tamper-evidence against attacks. We implement automated tools (on top of commercial CAD tools) for deploying guard wires around targeted nets within an open-source System-on-Chip.
Lastly, using an existing IC threat assessment toolchain, we show \dpar{} defeats even the stealthiest known hardware Trojan, with $\approx$\,1\% overhead.
\end{abstract}


\section{Introduction}\label{section:introduction}
Integrated circuits (ICs) are the foundation of computing
systems. Security vulnerabilities in silicon are devastating 
as they subvert even formally verified software. 
For almost 50 years, the transistors within ICs have continued 
to shrink, enhancing performance while reducing power and 
area usage. However, these advances that push the laws of 
physics come with a financial cost: the price to build a
3\,$nm$ fabrication facility capable of producing ICs at a 
commercial scale is estimated to be 
\$15--20B~\cite{cost_of_fab_3nm}. Even when
entities can afford to make such an investment, they must 
continually run the IC fabrication line (approximately 40,000 
wafers/month) as many fabrication processes cannot be readily 
stopped and restarted.

\begin{figure}[t]
\centering
\includegraphics[width=0.45\textwidth]{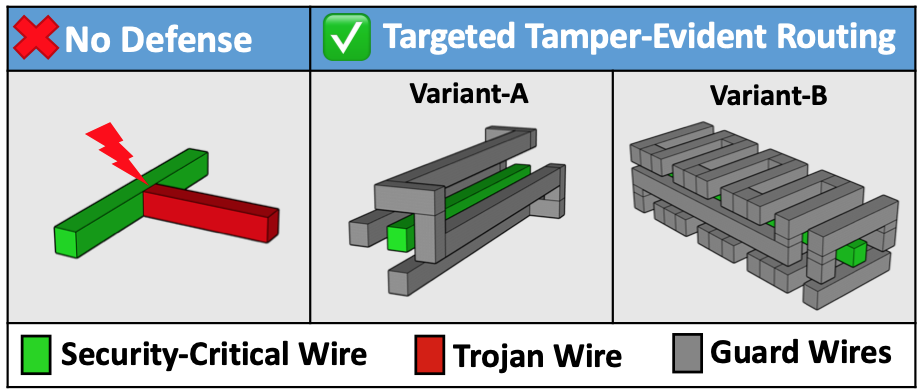}
\caption{\footnotesize \dpar{} is a \textit{preventive} layout-level defense against fabrication-time Trojans. \dpar{} deploys tamper-evident guard wires around security-critical wires in a circuit layout---in a pattern similar to variant A or B---to prevent attackers from attaching Trojan wires to them.}
\label{fig:def_routing_summary}
\figline{}
\vspace*{-0.2in}
\end{figure}

This extreme cost forces most semi-conductor companies, and even
nation states, to become ``fabless'', i.e., they outsource
fabrication. Today, only 3 companies in the world
(Intel, Samsung, and TSMC) have capabilities to fabricate ICs at the
10/7\,$nm$ process nodes~\cite{gf_7nm}. This presents a security threat:
fabless semiconductor companies and nation states must trust these
three manufacturers (and their partners) not to alter their designs at
any point throughout the fabrication process (i.e., implant a hardware Trojan).

The most stealthy and controllable hardware Trojans involve
inserting additional\footnote{Additive hardware Trojans are a class of
Trojan designs that require additional hardware to be added to a
circuit design. We are unaware of any documented stealthy and controllable subtractive or substitution Trojans. Dopant-level Trojans~\cite{kumar2014parametric,becker2013stealthy,shiyanovskii2010process} are the closest to such; however, they have limited controllability and are
detectable~\cite{sugawara2014reversing}.} 
circuit components designed to maliciously subvert the functionality of 
the chip (i.e., an additive hardware Trojan). Specifically, the A2 Trojan~\cite{a2} utilizes only two additional 
cells --- one analog  capacitor and one digital logic gate ---
to provide a hardware foothold~\cite{king08} within a microprocessor IC for 
an attacker to gain unauthorized supervisor privileges with user-mode code.

There are now only two ways of defending against hardware Trojans
implanted at fabrication-time: post-fabrication \textit{detection}
\cite{agrawal2007trojan,potkonjak2009hardware,jin2008hardware,zhou2015detecting,li2008speed,forte2013temperature} 
and pre-fabrication \textit{prevention}~\cite{xiao2013bisa,ba2016hardware}. 
The former tries to detect the presence of Trojan components after the
chip has been fabricated, while the latter attempts to alter the IC's
physical layout, at design time, in a way that makes foundry-side 
alterations challenging to an attacker. 

\textit{Detection} is more commonly studied than prevention and 
consists primarily of two techniques~\cite{tehranipoor2010survey}: 
1) side-channel analysis and 2) functional testing. 
Side-channel analysis attempts to detect noticeable deviations in 
power usage, electromagnetic (EM) emanations, performance (timing), etc.~\cite{agrawal2007trojan,jin2008hardware,potkonjak2009hardware,narasimhan2011tesr}. 
It often requires a ``golden'' reference chip to be effective, 
and can only detect the side-channel signature deviations greater than those
caused by process variation (i.e., the hardware Trojan must have a
large physical footprint). Alternatively, functional testing attempts
to inadvertently trigger the Trojan by activating as many logic paths
through the circuit as possible. Functional testing does not require
any ``golden'' reference chip, but it requires the Trojan's trigger to
be activated by the IC's common mode operation, as exhaustive testing
of even a moderately complex integrated circuit is infeasible.

Albeit less studied, \textit{prevention} is another defense against 
fabrication-time hardware Trojans. To prevent such attacks, we advocate that the \emph{placement and routing of security critical circuit elements should be a first-class part of an IC's back-end design}, on the level of performance, power, and cost.
To the best of our knowledge, 
only three preventive fabrication-time defenses have been
explored~\cite{xiao2013bisa,ba2015hardware,ba2016hardware}. 
All of them are \textit{placement}-centric, attempting to increase the device
layer (core) density by filling empty spaces with with tamper-evident 
logic gates, thus making it challenging for an 
attacker to find open space in the design to insert their Trojan components 
(cells/gates). However, there are several problems with placement-centric
defenses. As Ba {\em et al.}~\cite{ba2015hardware} point out, the
BISA cell approach~\cite{xiao2013bisa} is infeasible as it requires 
100\% placement density. Contrast this with
the 60-80\% density of current IC layouts that ensures
routability. If 100\% density were feasible, every IC design would be
manufactured that way to save cost. Alternatively, Ba {\em et
al.}~\cite{ba2015hardware,ba2016hardware} suggest targeted filling: only
filling placement sites that are located closest to ``security-critical'' logic. 
While prioritizing security-critical logic is a significant improvement, focusing on the device layer only impedes attacks due to inflated timing requirements, it does not prevent them, as \S\ref{subsection:eval_effectiveness_rd_results} shows.

Unfortunately, no single technique is effective in detecting, 
and/or preventing the insertion of the stealthiest known 
additive hardware Trojan, the A2 Trojan~\cite{a2}, which requires 
only two additional cells. To fill this gap, we propose \textit{\dparfirst{}}, 
a \textit{routing-centric} defense that
\textit{prevents} foundry-side attackers from routing Trojan wires
to, or directly adjacent to, security-critical wires. We define \dpar{}  
as any routing method that protects security-critical wires from fabrication-time
alterations. Specifically, we leverage concepts from the signal-integrity 
domain~\cite{hollis2006rasp,hollis2009pulse} and apply them to a security 
domain (addressing several technical challenges along the way): we route
``guard wires'' around security-critical wires that make it infeasible
for an attacker to tap any such wire without detection 
(i.e., tamper-evident), something characteristic of additive Trojans~\cite{icas}
(Fig.~\ref{fig:def_routing_summary}). Extending signal-integrity domain 
techniques to the security domain entails two technical challenges:
\begin{enumerate}
    \item \textit{completely} shielding all surfaces of critical wires, 
    \item and be tamper-evident.
\end{enumerate}
Contrary to placement-centric defenses, which focus on preventing attack 
\textit{implementation}, \dpar{} focuses on preventing attack 
\textit{integration}, and thus, does \emph{not} require filling 
\textit{all} the empty space in an IC design to be effective. 

We make the following contributions:
\begin{itemize}[nosep]
\item \dparfirst{}: a routing-centric, preventative, defense against stealthy IC 
    fabrication-time attacks. \dpar{} places \textit{tamper-evident} guard wires 
    alongside security-critical wires, making fabrication-time modifications to such 
    wires infeasible and/or detectable post-fabrication. 
\item Characterization of possible guard wire bypass attacks.
\item Attack-driven design of \textit{designed-in guard wires}. 
    Designed-in guard wires are added during the place-and-route 
    phase of the IC design process for the sole purpose of defending security-critical
    wires. They have minimal routing constraints and can guard all surfaces of 
    designer-targeted wires.
\item Automated routing toolchain for deploying guard wires within an IC layout that integrates with       commercial and open-source VLSI CAD tools.
\item Evaluation of the effectiveness of \dpar{} compared to previous defenses against both digital and analog A2 Trojans embedded in a System-on-Chip intended to be a surrogate for DoD systems of interest~\cite{cep}, using a recently published fabrication-time threat assessment tool~\cite{icas}.
    The results indicate \dpar{} is more effective than existing 
    placement-centric defenses~\cite{xiao2013bisa,ba2015hardware,ba2016hardware}, and 
    is capable of thwarting even the stealthiest additive hardware Trojans, including 
    A2~\cite{a2}.\footnote{It is important to note that routing-centric and placement-centric defenses are compatible (belt and suspenders). A designer would first apply \dpar{}, then fill open placement sites in a targeted manner.}
\end{itemize}

\section{Background}\label{section:background}
\subsection{IC Design Process}\label{subsection:ic_design_process}
\begin{figure}[t]
\centering
\includegraphics[width=0.45\textwidth]{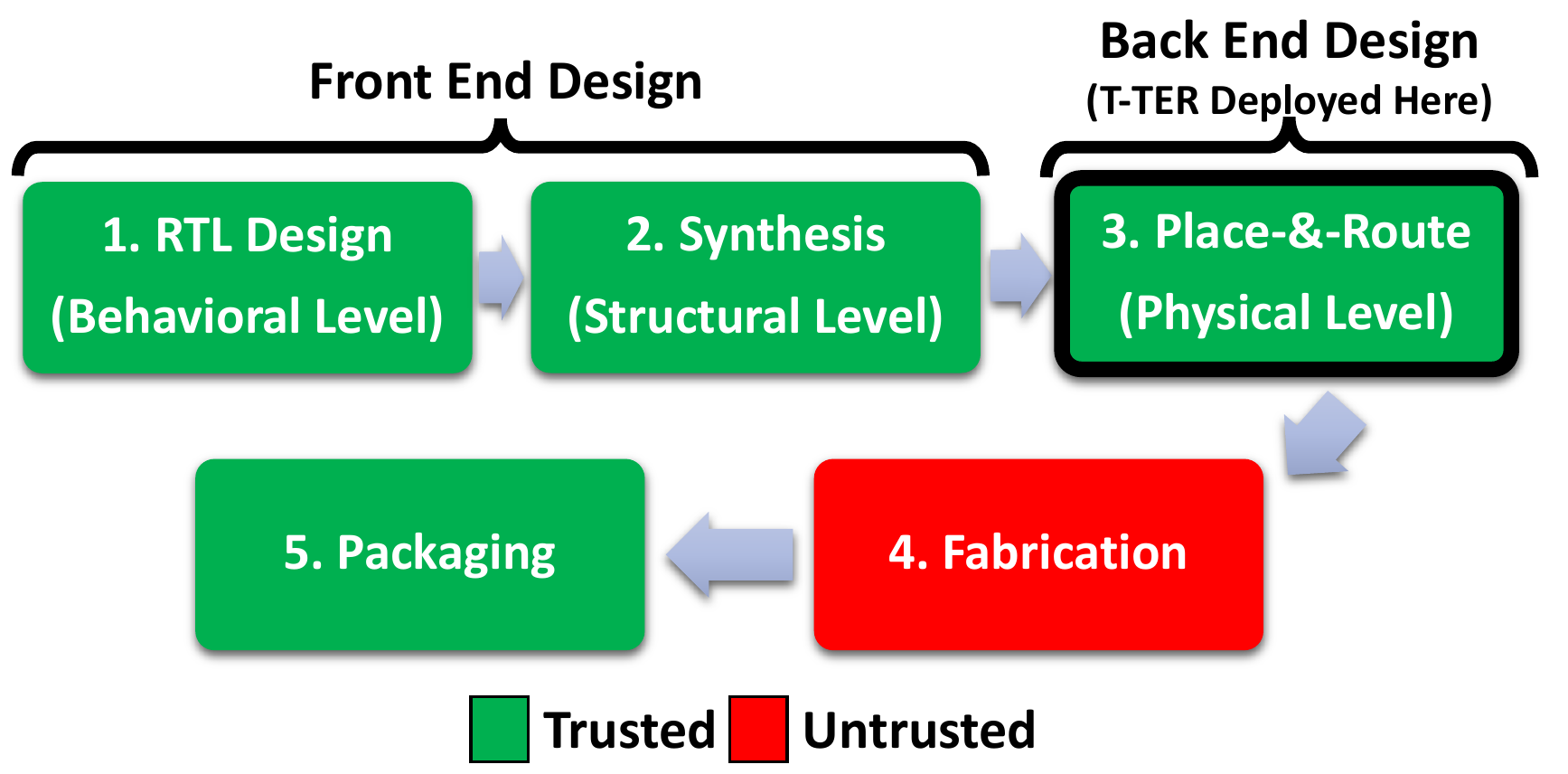}
\caption{\footnotesize The IC design process consists of five main phases. 
We assume fabrication (phase 4) is the only untrusted phase, as this is often 
outsourced due to economic forces. \dpar{} is deployed at the 
place-\&-route (layout) phase.}
\label{fig:ic_design_process}
\figline{}
\vspace*{-0.2in}
\end{figure}

Creating an Integrated Circuit (IC) consisting of a billion transistors is a complex process that 
requires its decomposition into sub-processes and extensive use of automation via Computer Aided 
Design (CAD) tools. The IC design process consists of five main phases, as illustrated in 
Fig.~\ref{fig:ic_design_process}. First, during RTL design, high-level descriptions of the IC are 
written in Hardware Description Languages (HDL) like Verilog or VHDL. Next, during synthesis, 
the HDL code is ``compiled'' into a gate-level netlist. The gate-level netlist is then 
placed-and-routed (PaR), and a physical geometric blueprint of the chip is encoded in a Graphics 
Database System II (GDSII) file. Lastly, the IC is fabricated, and packaged into a device for 
mounting on a printed circuit board. In line with prior work on untrusted foundry \cite{icas,a2,ba2015hardware,ba2016hardware,xiao2013bisa,kumar2014parametric,becker2013stealthy,lin2009trojan}, and economic forces, we assume all design phases---except fabrication---are trusted.

\begin{figure}[t]
\centering
\includegraphics[width=0.48\textwidth]{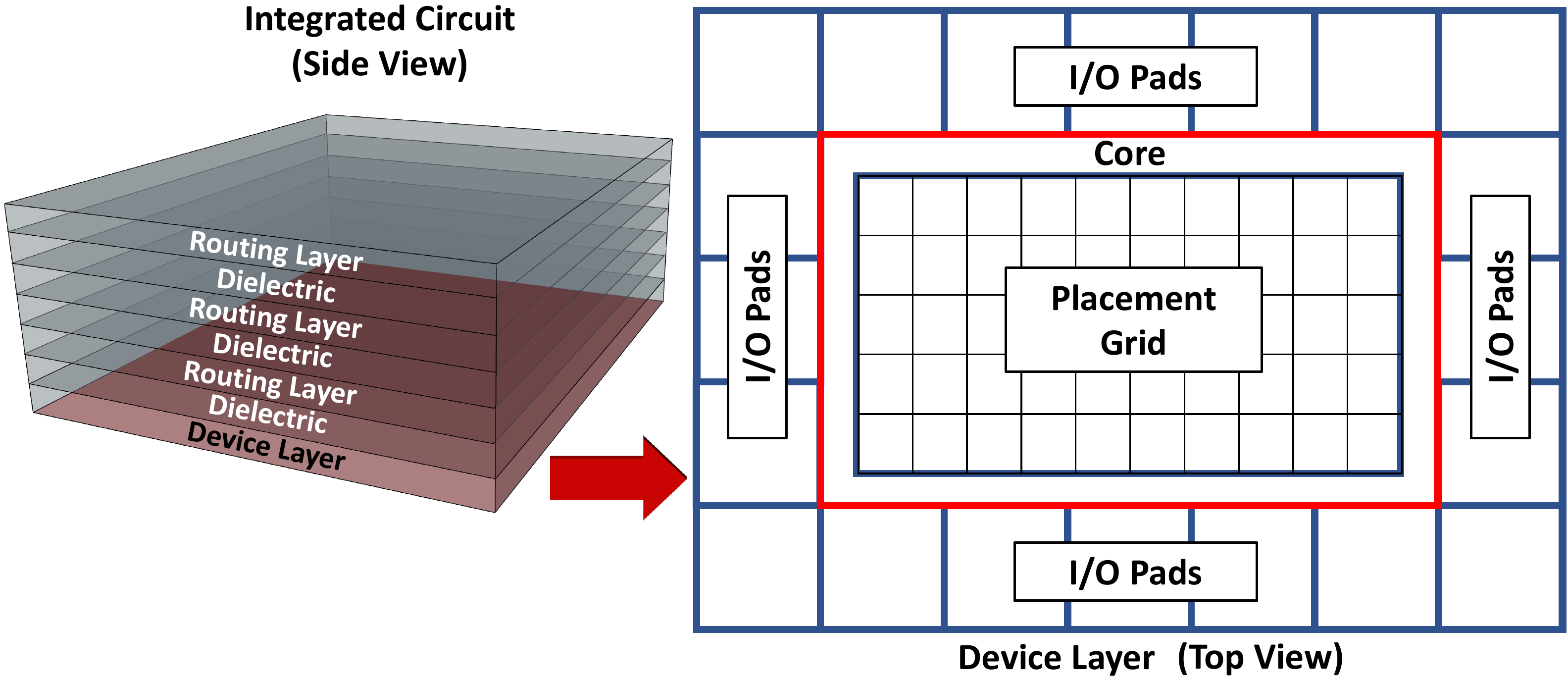}
\caption{\footnotesize Typical 3D physical IC layout designed during the place-and-route IC design 
phase (Fig.~\ref{fig:ic_design_process}). On the bottom is a device layer, and stacked above are 
several routing layers.}
\label{fig:ic_layout}
\figline{}
\vspace*{-0.1in}
\end{figure}

Defensive Routing is deployed at the physical level, i.e., the PaR design phase. 
During PaR, the gate-level netlist is physically arranged onto a 3-dimensional grid, 
shown in Fig.~\ref{fig:ic_layout}. The 3D grid consists of a device layer, where circuit 
components (e.g., digital logic gates) are placed, and several routing layers vertically 
stacked above, where wires are routed to connect the circuit components on the device layers. 
Each layer is separated by an insulating dielectric, and vias are used to 
connect wires on adjacent layers.

\subsection{Hardware Trojans}\label{subsection:hardware_trojans}

A hardware Trojan is a malicious modification to a circuit designed to 
alter its operative functionality~\cite{beaumont2011hardware}. 
It consists of two main building blocks: a \textbf{trigger} and \textbf{payload} 
\cite{icas,chakraborty2009hardware,jin2008hardware,wolff2008towards}. 
Prior work provides hardware Trojan taxonomies based on the type of 
trigger and payload designs they employ 
\cite{chakraborty2009hardware,jin2008hardware,wolff2008towards,icas}. Likewise, we adopt the same taxonomy.

\textbf{Trigger.}
The trigger is circuitry that initiates the delivery of the payload when it encounters a specific 
state. The goal of the trigger is to control payload deployment such that it is hidden from test 
cases (stealthy), but readily deployable by the attacker (controllable). Triggers are created 
by adding, removing, and/or manipulating existing circuit 
components~\cite{tehranipoor2010survey,a2,shiyanovskii2010process,kumar2014parametric}, and 
can be digital or analog~\cite{rostami2013hardware,a2,king08}. The ideal trigger---e.g., 
A2~\cite{a2}---achieves stealth and controllability while being small (i.e., requiring 
few additional circuit components).

\textbf{Payload.}
The payload is circuitry that, upon being signaled by the trigger, alters the functionality of the 
victim (host) circuit. Like the trigger, the payload can be analog or digital, and has a variety 
of possible malicious effects. Prior work demonstrates Trojan payloads that leak 
information~\cite{lin2009trojan}, alter the state of the IC~\cite{a2}, and render the IC 
inoperable~\cite{shiyanovskii2010process}. \emph{One attribute all documented controllable 
hardware Trojans have in common is that they must route a rogue wire to, or directly adjacent to, 
a security-critical wire within the victim IC~\cite{icas}.}

\textbf{Fabrication-Time Attacks}\label{subsection:fab_time_attacks} 
Inserting a hardware Trojan at fabrication time is different from inserting a Trojan 
during the front-end design. Unlike behavioral or structural-level attackers that maliciously 
modify the HDL or gate-level netlist, 
respectively~\cite{alkabani2008designer,jin2010dftt,waksman2013fanci}, 
the fabrication-time attacker only has access to the \textit{physical-level} representation 
of the IC design (i.e., output of phase 3 in Fig.~\ref{fig:ic_design_process}). 
Specifically, they must edit the geometric representation of the \textit{circuit layout}, e.g., 
the GDSII file.  While this is more challenging than editing the design at the behavioral- (HDL) or 
structural-level (netlist), where design specific semantics are more readily interpretable, 
it is even more difficult to defend. The post-fabrication defender receives a literal black box 
from the foundry. Comprehensively inspecting each fabricated die to verify the absence of 
malicious perturbations is infeasible for the most advanced hardware Trojans~\cite{a2}.

As previous research reveals, \textbf{implanting a hardware Trojan into an IC layout requires three steps~\cite{icas}: 1) \textit{Trojan Placement}, 2) \textit{Victim/Trojan Integration}, and 3) \textit{Intra-Trojan Routing}.}
\textit{Trojan Placement} is the process of finding empty space on the IC's device layer to 
add additional circuit components, e.g., logic gates, to construct the Trojan trigger 
and payload. \textit{Victim/Trojan Integration} 
requires attaching a rogue Trojan wire, or routing it directly adjacent, to an unblocked surface 
on a security-critical wire(s). Lastly, \textit{Intra-Trojan Routing} involves routing 
the Trojan circuit components to the Victim/Trojan integration 
point---the unblocked security-critical wire segment.    

\textbf{Layout-Level Defenses.}\label{subsection:layout_level_defenses}
Prior work attempts to thwart fabrication-time attacks by increasing the difficulty
of \textit{Trojan Placement}: filling empty space on the IC's device layer with 
temper-evident functional logic 
gates~\cite{xiao2013bisa,ba2015hardware,ba2016hardware}. 
As shown in 
\cite{icas}, this approach
is only effective for Trojans with large footprints, as filling all placement sites is 
infeasible~\cite{ba2015hardware}, and even targeting fill around security-critical logic~\cite{ba2016hardware} leaves the IC layout vulnerable to Trojans with small footprints~\cite{a2}. 
Orthogonally, \textit{\dpar{} targets \textit{Victim/Trojan Integration} by directing protection, at the routing level, around wires Trojans want to attach to.}

\subsection{Time-Domain Reflectometry (TDR)}\label{subsection:background_tdr}

Time-domain reflectometry (TDR) is an electrical analysis technique used to measure physical 
characteristics about a transmission line (i.e., a wire) such as length, number and distance between impedance discontinuities (e.g., bends), propagation delay, dielectric constant, 
etc.~\cite{hayden1994characterization,hsue1997reconstruction}.
Foundries already use TDR to perform root cause analysis on chips that fail post-fabrication testing---often during bring-up of a new process node. 
TDR works by characterizing a wire within a circuit by injecting a 
single rising pulse down that wire and analyzing its reflection(s).

\textbf{IC Interconnect Models.}
There are two ways to model IC interconnects: lumped and transmission-line 
models~\cite{bakoglu1990circuits}. Lumped interconnect models approximate interconnects using 
networks of resistors and capacitors. Transmission-line models approximate interconnects 
as transmission lines with a characteristic impedance and propagation delay.

The choice of interconnect model is a function of maximum frequency component to wire 
length~\cite{sutherland1999edge}. A common rule of thumb for IC interconnects is: \textit{a wire 
is considered a transmission line if its length is greater than $\approx$10\% of the wavelength of 
the maximum frequency component it transmits~\cite{sutherland1999edge}.} In digital electronics, 
it is common to think of signals in terms of rise and fall times, rather than maximum frequency 
component. Thus, one can modify the prior rule of thumb to: \textit{a wire is considered as a 
transmission line if the transmitted signal rise time, $T_{rise}$, is less than twice the wire's 
propagation delay, $T_{pd}$~\cite{sutherland1999edge}}. Eq.~(\ref{eq:transmission_line_model}) 
captures this rule of thumb. 

\vspace*{-0.2in}
\begin{equation}
\label{eq:transmission_line_model}
    \text{Model} = 
\begin{cases}
    \text{Transmission Line},&  T_{rise}< 2T_{pd}\\
    \text{Lumped RC},        & \text{otherwise}
\end{cases}
\end{equation}

Choosing the right model is vital to understanding operational limitations and ensuring signal 
integrity within an IC layout. For example, an interconnect that carries a high-speed signal 
transitions will observe signal reflections from impedance discontinuities that are destructive to 
the signal integrity of the overall system. Modeling such interconnects using a lumped RC model 
can hide these destructive effects, while a transmission-line model would not.

\textbf{TDR for IC Fault Analysis.}
By Eq.~(\ref{eq:transmission_line_model}), the faster the rising edge of TDR's incident pulse, 
the finer-grain of propagation delay changes are detectable. 
TDR was first developed as a fault-analysis technique for long transmission lines, 
such as telephone or optical communication lines~\cite{somlo1969microwave,philen1982single}. 
As commercial TDR systems became more advanced, 
TDR became a standard IC packaging fault analysis 
tool~\cite{odegard1999comparative,smolyansky2004electronic,chen2006nondestructive}. 
Researchers have now demonstrated terahertz- level TDR systems capable of locating faults 
in IC interconnects to nanometer-scale accuracies
\cite{nagel2011contact,tay2012advanced,cai2010electro,teraview}. 
With such fine-grain resolution, \textbf{TDR is an ideal tamper-analysis technique for ensuring the 
integrity of the guard wires} used in \dpar{} (\S\ref{subsection:eval_threat_analysis}).

\section{Threat Model}\label{section:threat_model}
We adopt a threat model in which all phases of the IC design process are trusted 
\textit{except} fabrication (Fig.~\ref{fig:ic_design_process}). 
The untrusted foundry threat model stems from the extreme ramp-up costs associated with 
fabricating leading-edge silicon~\cite{cost_of_fab_3nm,gf_7nm} that make outsourcing IC 
fabrication a necessity---even for nation states. In line with previous untrusted foundry threat 
models~\cite{icas,a2,rostami2013hardware,tehranipoor2010survey,lin2009trojan}, we assume the 
worst case: that any fabrication-time modifications are carried out by a malicious actor 
within the foundry (or any foundry partners) that has access to the entire physical layout 
of the IC in the form of a GDSII file. 

While there are many types of hardware Trojans~\cite{rostami2013hardware} 
(\S\ref{subsection:hardware_trojans}), we focus on additive Trojans, rather than subtractive 
or substitution Trojans. Additive Trojans require implanting additional circuit components and 
wiring into the IC design.
We focus on additive Trojans as there are no documented stealthy and controllable examples of subtractive or substitution Trojans that we are aware of. The closest example of such Trojans are
dopant-level Trojans~\cite{shiyanovskii2010process,kumar2014parametric,becker2013stealthy}, 
all of which have limited controllability and are detectable with optical 
microscopy~\cite{sugawara2014reversing}.

Previous work shows that to successfully implement an \textit{additive} hardware Trojan, the adversary must complete the 
three steps---\textit{Trojan Placement}, \textit{Victim/Trojan Integration}, and 
\textit{Intra-Trojan Routing}~\cite{icas}---without being exposed. Namely, they must 1) find 
empty space on the device layer to insert the Trojan's components (logic gates/cells), 
2) locate an unblocked segment on a security-critical wire to attach the Trojan to, 
and 3) route the Trojan components to that unblocked wire segment. 
They are restricted from modifying the dimensions of the chip and/or violating 
manufacturing design rules that would risk their exposure. They are allowed to move components 
and/or existing wiring around, but are constrained by available resources (e.g., time) 
and correctness from making mass perturbations to the layout. 
As process technologies scale, manufacturing design rules become increasingly 
complex~\cite{dr_complexity_rising}. Thus, rearranging components and/or existing wiring comes 
at a substantial cost. The time to complete any layout modifications, and verify such 
modifications have not violated design correctness, cannot disrupt the fabrication turn-around 
time expected by their customers.\footnote{Typically, fabrication turn-around times are 
$\approx$3 months~\cite{tsmc_turnaround_time,battling_fab_cycle_time}.} 
Additionally, the attacker avoids any modifications that are detectable using existing 
test-case or side-channel based defenses. While it would be trivial for an attacker with 
\textit{infinite} time and resources to reverse-engineer the physical layout into HDL, 
add a Trojan, and re-run the design through the entire IC design process 
(Fig.~\ref{fig:ic_design_process}) thus generating an entirely new layout,
such an attack will be infeasible within the hard time limits of fabrication contracts, 
thus outside the scope of our threat model.

\section{\dparfirst{}}\label{section:guard_wires}
\dpar{} aims to make the second step of Trojan insertion---\textit{Victim/Trojan Integration} 
(\S\ref{subsection:fab_time_attacks})---intractable by shielding the surfaces of targeted 
wires (interconnects) with tamper-evident guard wires (\S\ref{subsection:layout_level_defenses}), 
creating an additional obstacle for adversaries to overcome. Similar to prior 
work~\cite{icas,ba2015hardware,ba2016hardware,linscott2018swan}, \dpar{} is made  practical by 
leveraging the observation that, for most hardware designs, only a subset of the IC is security-critical~\cite{specs15,zhang2017identifying,jin2010dftt,glift,zhang2020transys,linscott2018swan}, or the target of a hardware Trojan. In designing \dpar{}, we pose three questions:
\begin{enumerate}[nosep]
    \item \textit{Which wires in the design are security-critical (should be guarded)?}
    \item \textit{How can an attacker bypass \dpar{} guard wires?}
    \item \textit{How do we design guard wires that are tamper-evident with respect to bypass attacks?}
\end{enumerate}

\subsection{Identifying Security-Critical Nets}\label{subsection:iding_sc_wires}
While identifying security-critical features, and corresponding nets (wires), in a design 
is an orthogonal problem---and an ongoing area of research~\cite{specs15,zhang2017identifying,jin2010dftt,glift,zhang2020transys,linscott2018swan}---%
selecting  wires to guard is the first step in deploying \dpar{}. 
Currently, there exist two techniques for identifying security-critical nets: 
1) \textit{manual}~\cite{specs15,jin2010dftt,linscott2018swan} or 
2) \textit{semi-autonomous}~\cite{zhang2017identifying,zhang2020transys}. 
In \textit{manual} identification, a human expert analyzes the design's specification, and 
the corresponding HDL, and flags nets that implement features critical to the 
security of software 
or other hardware that interface to the design~\cite{specs15,jin2010dftt,linscott2018swan}. 
Alternatively, in \textit{semi-autonomous} identification, a set of security-critical nets for a 
specific design are first manually identified~\cite{specs15,jin2010dftt}, or mined from a list 
of published errata~\cite{zhang2017identifying}, and either: 1) used to train a classifier that 
identifies similar nets in other designs~\cite{zhang2017identifying}, 2) expanded using 
information flow~\cite{glift} or fan-in analyses~\cite{icas}, or 3) translated to an entirely 
different design~\cite{zhang2020transys}.
In this paper, we adopt the most common approach in this area of semi-autonomous identification~\cite{ba2016hardware,icas}.

\subsection{Guard Wire Bypass Attacks}\label{subsection:gw_bypass_attacks}
\begin{figure}[t]
\centering
\includegraphics[width=0.45\textwidth]{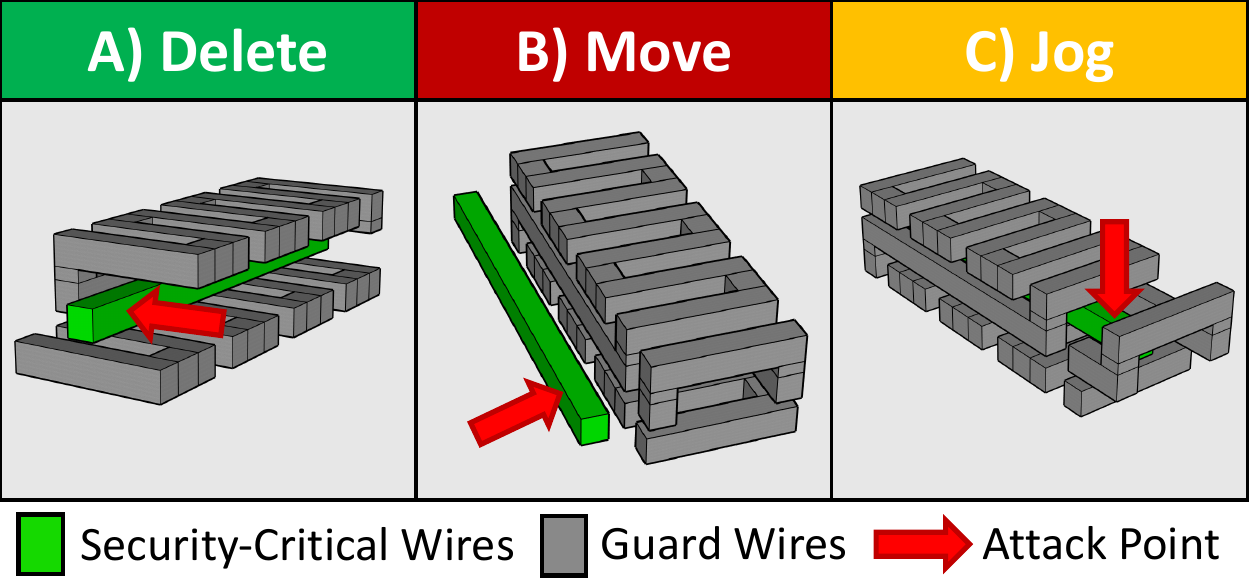}
\caption{\footnotesize There are three ways an attacker could bypass \dpar{} guard wires to 
connect a Trojan wire to a security-critical wire, color-coded by attacker difficulty: 
A) \textit{delete} guard wire(s), B) \textit{move} an intact set of guard wires, 
or C) \textit{jog} guard wires out of the way. We study the \textit{jog} attack to 
assess defensive sensitivity, as it strikes a balance in attacker difficulty, and 
is the most difficult to detect.}
\label{fig:gw_attacks}
\figline{}
\vspace*{-0.2in}
\end{figure}

With \dpar{} deployed, attackers must \textit{bypass} guard wires---by exposing the surface of a 
security-critical wire(s)---to complete \textit{Victim/Trojan Integration}, i.e., connect a rogue 
Trojan wire to a security-critical wire(s) (\S\ref{subsection:fab_time_attacks}). 
Given a set of interconnected guard wires (Fig.~\ref{fig:def_routing_summary}), there are 
three ways an attacker can bypass them, color-coded by attacker difficulty 
(Fig.~\ref{fig:gw_attacks}): A) delete, B) move, or C) jog attacks. In a \textit{deletion} attack 
(Fig.~\ref{fig:gw_attacks}A), entire guard wire(s) are removed from the layout. 
While this attack is easy to implement, it is also easy to defend. A post-fabrication continuity 
check of a connected set of guard wires will detect a deletion attack. 
In a \textit{move} attack (Fig.~\ref{fig:gw_attacks}B), all interconnected guard wires 
are left intact, but translated to another location on the chip. 
Move attacks are the most difficult to implement: an attacker must find a contiguous group 
of unused routing tracks to translate each set of guard wires too. 
Even then, a post-fabrication cross-talk analysis between security-critical and guard wires 
would expose this attack~\cite{potkonjak2009hardware,hollis2006rasp}. 
Lastly, in a \textit{jog} attack, guard wires are \textit{lengthened} to make room for 
a rogue Trojan wire to connect to a security-critical wire using a via. 
Jog attacks strike a compromise in terms of implementation difficulty, and are the stealthiest 
of all bypass attacks. They are easier to implement than move attacks, and are undetectable 
with post-fabrication continuity tests or cross-talk analyses. \textit{The only artifacts of a 
jog attack are: 1) a change in the number of bends in the guard wire, i.e. number of impedance discontinuities, and/or 2) an increase in the guard wire's length.} However, nanometer scale time-domain reflectometry (TDR)~\cite{tay2012advanced,nagel2011contact} detects these changes (\S\ref{subsection:eval_threat_analysis}).

\subsection{Tamper-Evident Guard Wires}\label{subsection:gw_types}

While techniques for detecting all three bypass attacks exist, each of them requires the ability 
to measure physical characteristics (e.g., continuity, cross-talk, and length) about a guard 
wire post-fabrication. \textit{How do we design guard wires whose physical characteristics are 
tamper-evident post-fabrication?} Based on these considerations, we take a straw-man approach in  designing guard wires capable of preventing even the stealthiest of attacks. 


\subsubsection{\textbf{Na\"ive Approach: Re-purpose Existing Wires}}\label{subsection:defensive_routing_natural}
One idea for constructing guard wires is to re-purpose existing
non-security-critical wires, inherent to the host IC design, as guard wires. 
Such an approach creates hyper-local routing densities nearby security-critical 
wires, thus limiting or eliminating the locations where an attacker 
can attach rogue Trojan wires. By re-purposing pre-existing wires as guard wires, the guard 
wires incur no hardware overhead. Unfortunately, there are additional 
routing constraints (e.g., toggle frequency, length, layer, location, timing sensitive, and spacing) that limit the pool of candidate guard wires.
Even when such constraints are met, the guard wires are only tamper-evident with respect to deletion and move attacks.
For an existing wire to also be tamper-evident with respect to the more stealthy jog and bypass attacks, it must be timing-critical (i.e., if it is made longer, then it will cause timing violations that manifest as run-time errors).
As Fig.~\ref{fig:gwire_type_nb} shows, deployment using existing guard wires is challenging.
Namely, the lack of suitable wires in many designs makes it infeasible to block all surfaces of all security-critical wires.

\subsubsection{\textbf{Designed-in Guard Wires}}\label{subsection:defensive_routing_synthetic}
To fill the gaps of existing wires, we propose designed-in guard wires.
Designed-in guard wires are \textbf{not} inherent to the host IC design. Rather, 
they are added to the design during the place-and-route IC design phase 
(Fig.~\ref{fig:ic_design_process}). Since they do not implement any circuit functionality, they
have fewer routing constraints. As we show in Fig.~\ref{fig:gwire_type_nb}, completely blocking the accessible surface area of all security-critical wires is trivial. While designed-in guard wires incur hardware overhead, i.e., additional wires, they completely block an attacker from attaching a Trojan wire at fabrication time (Victim/Trojan Integration, \S\ref{subsection:fab_time_attacks}), as shown in Fig.~\ref{fig:gwire_type_rd}. Additionally, designed-in guard wires are tamper-evident with respect to \textbf{\textit{all}} bypass attacks, when coupled with post-fabrication analysis techniques like continuity checking, cross-talk analysis, and time-domain reflectometry (\S\ref{subsection:background_tdr} and 
\S\ref{subsection:eval_threat_analysis}), repsectively.

There are several designed-in guard wire architectures that may be deployed, listed in order of 
increasing difficulty of deployment: 1) fully-disjoint, 2) partially-connected, and 3) 
fully-connected. Fully-disjoint designed-in guard wires are not connected between sides, i.e., 
the guard wires on each side of a security-critical wire are never connected to one another. 
Partially-connected guard wires allow for a single guard wire to be utilized on multiple sides. 
For example, a security-critical wire could be guarded on the north, east, and west sides by a 
single guard wire that wraps around the security-critical wire. Lastly, fully-connected guard 
wires are formed when a single guard wire is routed around all sides of all security-critical 
wires, as shown in Fig.~\ref{fig:def_routing_summary}. 

To detect tampering of designed-in guard wires post-fabrication, their analog characteristics of
must be observable. This can be implemented either on-chip, e.g., with internal 
sensors~\cite{kelly2015detecting} or ring oscillators~\cite{zhang2011ron}, or off-chip, e.g., 
with two I/O pins and a one-time programmable fabric~\cite{linscott2018swan}. If fully-joint or 
partially-connected designed-in guard-wires are deployed, the one-time programmable fabric could 
be randomly programmed to route both ends of a single (fully-disjoint) or single-set 
(partially-connected) of guard wire(s) to the two pins. If fully-connected designed-in guard 
wires are deployed, the one-time programmable fabric is not needed, as both ends of the guard 
wires set can be routed to the two pins.


\section{Implementation}\label{section:implementation}
We develop an automated toolchain for deploying \dpar{} in modern IC designs. Our toolchain 
integrates with existing IC design flows (Fig.~\ref{fig:ic_design_process}) that utilize 
commercial VLSI CAD tools. Specifically, we implement the \dpar{} toolchain around the Cadence 
Innovus Implementation System~\cite{innovus}, a commercial place-and-route (PaR) CAD tool. 
The toolchain is invoked by modifying a place-and-route TCL script,\footnote{Tool Command Language (TCL) scripts are the standard programmatic interface to commercial VLSI CAD tools. IC designers often develop a set of scripts for driving the CAD tools that automate most of the IC design process (Fig.~\ref{fig:ic_design_process}).} as shown in Fig.~\ref{fig:def_routing_toolchain}.

\subsection{Place-\&-Route Process}\label{subsection:implementation_par}
The PaR design phase (Fig.~\ref{fig:ic_design_process}) is typically automated by a CAD tool, 
programmatically driven by TCL script(s). There are several steps to PaR that are performed in 
the following order: 1) floor-planning, 2) placement, 3) clock tree synthesis, 4) routing, and 
5) filling. \textit{To ensure that all guard-wires are routed optimally, we modify the order of 
these PaR steps}. Specifically, after floor-planning (1), we use our automated toolchain to 
place identified components \textit{and} route identified wires and their guard wires. Our 
toolchain then permanently fixes the locations of these components and wires to prevent the PaR 
CAD tool from modifying their positions and/or shapes throughout the remainder of the PaR process. Lastly, we utilize the PaR CAD tool to place all other  components (2), synthesize 
the clock tree (3), route remaining wires(4) and fill the design with filler 
(capacitor) cells.

\subsection{Automated Toolchain}\label{subsection:implementation_toolchain}

\begin{figure}[t]
\centering
\includegraphics[width=0.4\textwidth]{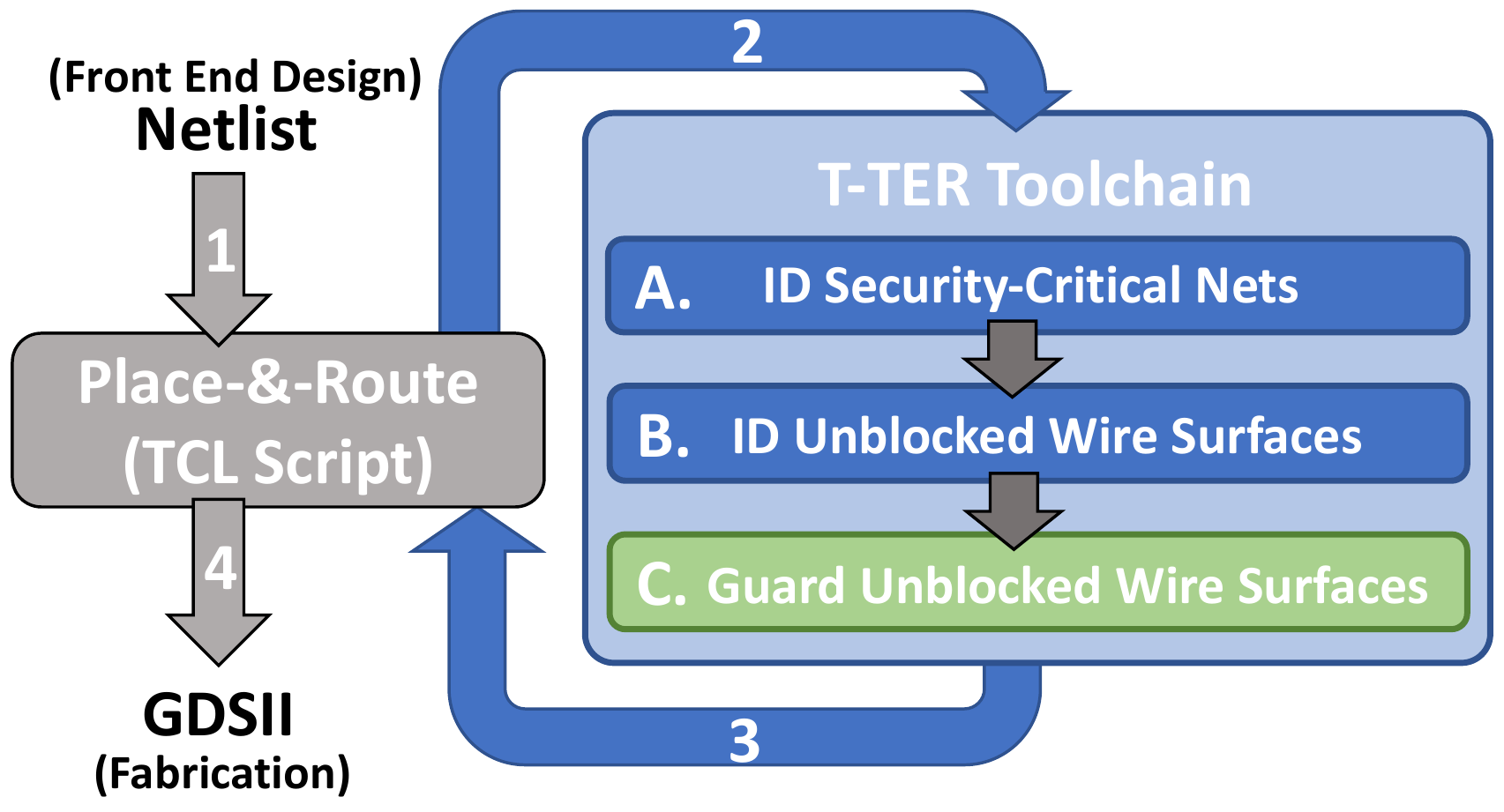}
\caption{\footnotesize \dpar{} is an automated toolchain consisting of three phases. Our toolchain 
first identifies which wires are security-critical, determines potential (unblocked) attachment points, 
and routes guard wires to block all attachment points. Identified components \& wires are placed \& routed \textit{before} phase (A) of our toolchain is invoked. Before continuing with the traditional 
PaR flow, the protected nets and their guard wires are locked in-place to ensure they are 
untouched throughout the remainder of the layout process.}
\label{fig:def_routing_toolchain}
\figline{}
\vspace*{-0.2in}
\end{figure}

The \dpar{} toolchain automates the insertion of either existing or designed-in guard wires around 
wires in need of protection. The toolchain consists of three main phases 
(Fig.~\ref{fig:def_routing_toolchain}). The first phase (A) identifies security-critical nets. 
The second phase (B) identifies the unblocked surfaces of all of these nets within 
a GDSII-encoded layout. The last phase (C) guards the nets and their influencer nets by 
routing guard wires nearby. We provide additional implementation details on all 
three stages of the \dpar{} toolchain below.

\textbf{Identifying Nets.}
\label{subsection:implementation_id_sc_nets}
The first phase of \dpar{} requires identifying nets in the design to guard, i.e., nets that are 
security-critical. Phase A of our toolchain (Fig.~\ref{fig:def_routing_toolchain}A) utilizes a 
semi-autonomous approach to identifying such nets (\S\ref{subsection:iding_sc_wires}). 
Specifically, our toolchain assumes the designer has \textit{manually} flagged a set of \textit{root} 
security-critical nets in the behavioral-level HDL by appending a unique prefix---\textit{secure\_}%
---to each signal (net) name. During PaR, our toolchain performs a data-flow analysis of 
the circuit netlist to locate the direct fan-in---to a configurable depth---of each root net.
Since the netlist is often modified by PaR CAD tools to meet various design 
constraints (e.g., power, performance, and area), we disable the optimization of all root
nets during PaR. Given the interconnected nature of nets within an IC design, an adversary may elect to 
target a net that influences a root net, rather than the root net itself. 
Our toolchain addresses this indirection, using an autonomous approach that widens the set of \textit{targeted} nets to the root nets and those that influence root nets (to a designer configurable degree).
The remainder of our tool flow focuses on protecting this set of targeted nets.

Our fan-in analysis tool is a custom-backend to the Icarus Verilog (IVL) front-end 
Verilog compiler~\cite{icarus}, and is implemented in C++. 
It performs a breadth-first search over the circuit-level data-flow graph 
generated by IVL. We release our fan-in analysis tool under an open-source license.

\textbf{Identifying Unblocked Wire Surfaces.}
\label{subsection:implementation_locate_sc_nets}
The second phase of \dpar{} is identifying the unblocked surfaces of targeted nets in a physical IC layout, i.e., potential locations of Trojan wire attachment. 
To do so, we implement, and open-source, a Python tool that analyzes the GDSII layout 
file containing only the placed-and-routed targeted components and wires. 
Our tool implements a 3-D scanning window approach to search the 3-D boundary 
surrounding each targeted wire, and compute the areas on each wire's 
surfaces that are not blocked by other wires or circuit components. 
While it is traditional for designers to only route wires on defined \textit{routing tracks}, 
i.e., on a pre-defined routing grid, it may be possible for an attacker to route 
Trojan wires off this grid, so long as they maintain the minimum spacing requirements 
dictated by the manufacturing design rules. Thus, our tool takes a conservative 
approach when scanning for unblocked wire surfaces, only scanning the 3-D boundary 
surrounding each targeted wire that extends up to the minimum-spacing 
requirements defined for the given, and adjacent (top/bottom), routing layers. 
If and only if another component or wire overlaps a region of the 3-D boundary surrounding 
a targeted wire, that surface region will be considered blocked. 
The output of this stage of our toolchain is a list of coordinates within the 3-D 
place-and-route grid that must be filled with guard wires during the next phase 
in the \dpar{} toolchain. 


\textbf{Guard Unblocked Wire Surfaces.}
\label{subsection:implementation_defensive_routing}
The last stage of the \dpar{} toolchain (Fig.~\ref{fig:def_routing_toolchain}) is 
a custom guard wire routing tool, also implemented in Python. 
It takes as input exact locations of targeted wires and their unblocked sides 
(output from Phase B, \S\ref{subsection:implementation_locate_sc_nets}) and generates 
a TCL script that integrates with the Cadence Innovus Digital Implementation 
platform~\cite{innovus} to automatically route the guard wires. 
This TCL script is executed immediately after the targeted wires have been 
routed, but before placing the remaining components. 
Depending on the guard wires being deployed, existing or designed-in,
different guard wire TCL scripts are generated (described below).\footnote{While existing guard wires fail to defend against all types of guard wire attacks (\S\ref{subsection:defensive_routing_natural}), we implement a tool to deploy them in order to empirically show they are also inferior to designed-in guard wires in terms of surface-are coverage (Figs.~\ref{fig:gwire_type_nb} \&~\ref{fig:gwire_type_rd}), and thus should not be used in a security context.}

There are numerous ways \textit{existing guard wires} can be implemented. 
Since commercial PaR CAD tools do not offer an interface to enable fine-grain 
constraints between two unrelated signal wires, we develop an indirect method 
for implementing existing guard wires. We implement existing guard wires by 
constraining placement and routing resources nearby targeted wires. 
First, we identify all circuit components (i.e., logic gates) connected to all 
targeted wires, i.e., targeted components. Next, we draw a bounding 
box around these components and extend this boundary vertically by 
\textit{Y}\% of the overall box height, and horizontally by \textit{X}\% of the overall box width. 
Then, we set placement and routing density screens in the portion of the IC layout that lies 
\textit{outside} the bounding box. These constraints limit the placement 
and routing resources outside the bounding box, thus forcing more components and 
wiring within the bounding box. 
With increased routing density nearby targeted wires, they are less accessible by 
Trojan payload delivery wires. The values of \textit{X}, \textit{Y}, and density 
screen configuration settings 
are optimized to maximize the net blockage metric computed by the GDS2Score metric. 

\textit{Designed-in guard wires} are more straightforward to implement. The automated guard wire 
deployment toolchain locates all unblocked surfaces (north, south, east, west, top, 
and bottom) of all targeted wires and routes guard wires in these regions. 
After all guard wire segments are routed, they are connected according to the architecture 
chosen (\S\ref{subsection:defensive_routing_synthetic}). 

\section{Evaluation}\label{section:evaluation}
We evaluate \dpar{} in three areas. First, we explore the effectiveness of \dpar{} at closing the 
fabrication-time attack surface of three security-critical features within an open-source 
System-on-Chip (SoC), with regard to the stealthiest additive Trojan known: the A2 Trojan~\cite{a2}. 
We compare the capabilities of \dpar{} with existing state-of-the-art layout-level 
defenses~\cite{xiao2013bisa,ba2015hardware,ba2016hardware}. Next, we demonstrate the practicality of 
\dpar{}, analyzing its power, performance, and area overheads. Finally, we perform a threat 
assessment, demonstrating how guard wires are tamper-evident.

\subsection{Experimental Setup}
\label{section:exp_setup}

\textbf{Surrogate SoC.}
We utilize the open-source Common Evaluation Platform (CEP) SoC 
design~\cite{cep_v1} for our evaluation. The CEP platform is designed as a surrogate SoC system for 
testing a variety of DoD-oriented IC technologies. It contains a general-purpose processor core, five cryptographic 
cores, four digital signal processing cores, and a GPS core. We focus on three cores from in the SoC: the \textit{processor} core, the \textit{DFT} core, and the \textit{AES} core. The OR1200 processor\footnote{We use the OR1200 version of the CEP rather RISC-V version since the OR1200 is the processor used in the A2 Trojan~\cite{a2}. We are not aware of similar Trojans available in the RISC-V. We expect similar results for the RISC-V version of the CEP since both processors are RISC-based, in-order, scalar, pipelined, capable of running Linux, and operate at similar clock frequencies. Thus, from an IC layout perspective, they have similar features (e.g., wire lengths) and will have similar hardware overheads.} is a 5-stage pipelined CPU that implements a 32-bit OR1K instruction set and Wishbone bus interface~\cite{or1200}, and is the same design used in previous fabrication-time attack studies~\cite{icas,a2}. It supports Linux via BusyBox~\cite{busybox}. The AES core supports 128-bit key sizes. The DFT accelerator implements a Discrete Fourier Transform algorithm, a common component of radar and other sensing systems. 

We target a 45\,$nm$ Silicon-On-Insulator (SOI) process technology, and synthesize our design with 
Cadence Genus (v16.23), and placed-and-route it using Cadence Innovus (v17.1). All layout variations 
of our SoC target a 100\,$MHz$ clock frequency and a core density of 60--80\%. All CAD tools are run 
on a server with 2.5\,$GHz$ Intel Xeon E5-2640 CPU and 64GB of memory, running Red Hat Enterprise 
Linux (v6.9).

\textbf{A2 Trojan.}
The goal of \dpar{} is to protect security-critical features within SoCs from the stealthiest additive 
Trojan currently known, the A2 Trojan~\cite{a2}. The A2 Trojan is stealthy, i.e., evades current \textit{prevention} and \textit{detection} defenses, due to its small size and complex triggering 
mechanism. When implemented within our surrogate SoC, in a 45\,$nm$ process, the analog variant of the 
A2 Trojan~\cite{a2} requires only two additional cells that occupy 20 placements sites, while the entirely 
digital variant of the same attack requires 91 additional cells that occupy 1,444 placement sites. 
The analog A2 attack is \textit{not} timing critical: the Trojan components may be placed anywhere on the 
placement grid, at any distance from the Victim/Trojan integration point. Conversely, the digital A2 
attack \textit{is} timing-critical: the length of the interconnect between the Trojan components and the 
Victim/Trojan integration point must be within three standard deviations from the mean net length in the 
overall SoC (this is an entirely worst-case estimate borrowed from 
\cite{icas}). We summarize the placement and routing resource requirements for the two variants of 
the A2 Trojan in Table~\ref{table:a2_trojan_designs}.

\begin{table}
\centering
\caption{\footnotesize A2 Trojans used in \dpar{} effectiveness assessment.}
\begin{tabular}{l c c c}
\multicolumn{1}{c}{\textbf{Trojan}} & 
\multicolumn{1}{c}{\textbf{\begin{tabular}[c]{@{}c@{}}\# Std\\ Cells\end{tabular}}}& 
\multicolumn{1}{c}{\textbf{\begin{tabular}[c]{@{}c@{}}\# Placement\\Sites\end{tabular}}}&
\multicolumn{1}{c}{\textbf{\begin{tabular}[c]{@{}c@{}}Timing\\Critical?\end{tabular}}}
\\ \hline \hline
\rowcolor{GreenHLight!30}A2 Analog~\cite{a2} &   2 &   20 & \xmark \\
\rowcolor{YellowHLight!30}A2 Digital~\cite{a2} &  91 & 1444 & \cmark \\
\end{tabular}
\label{table:a2_trojan_designs}
\vspace*{-0.2in}
\end{table}

\textbf{Exemplar Nets of Interest.}
For this evaluation, we need to protect nets that our example Trojan might want to use as integration points.
Leveraging existing hardware Trojan payloads, we select three reference integration targets within 
our SoC design to protect with \dpar{}:
\begin{enumerate}[noitemsep,nolistsep]
    \item processor supervisor bit (supv),
    \item DFT computation ready interrupt (next\_out),
    \item cryptographic key bits (key [0:127]).
\end{enumerate}
The most popular hardware Trojans leverage the supervisor (supv) net as part of privilege escalation attacks~\cite{king08,hicks10,a2}. 
Alternatively, hardware Trojans can also hide specific computations or state transitions, e.g., a Trojan that disables the DFT computation-ready interrupt signal (or \textit{next\_out} signal) that informs the CPU when a DFT computation is ready.
Lastly, another popular hardware Trojan seeks to leak cryptographic key bits via side channels~\cite{lin2009trojan}.
The A2 trigger can be attached to any of the nets that carry these signals to mount an attack, so we protect the interconnects that comprise these nets.

\begin{figure*}[h]
\centering
\includegraphics[width=\textwidth]{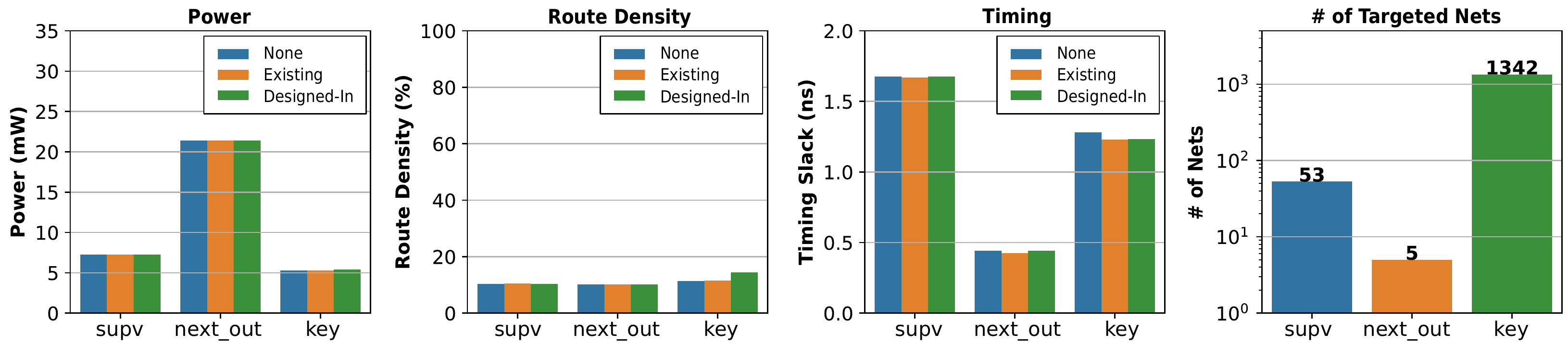}
\vspace*{-0.3in}
\caption{\footnotesize \dpar{} hardware overheads. The far right plot shows the number of wire 
(route) segments that implement the labeled security-critical feature (set of nets) in our surrogate SoC.}
\label{fig:defensive_routing_stats}
\figline{}
\vspace*{-0.2in}
\end{figure*}

The initial stage (Fig.~\ref{fig:def_routing_toolchain}A) of our automated \dpar{} toolchain assumes 
the designer has manually annotated the root nets they have chosen to target with 
\dpar{} (\S\ref{subsection:implementation_id_sc_nets}). Thus, we manually annotate the above net (signal) 
definitions with the prefix \textit{secure\_} within our SoC design's RTL. We then synthesize and 
place-and-route our design prior to generating a final, optimized, netlist for which our toolchain 
computes the fan-in to each manually annotated net---to a depth of two layers of logic gates---thereby 
expanding the final set of all targeted nets (i.e., those guarded by \dpar{}). 
Fig.~\ref{fig:defensive_routing_stats} (far right) shows the number of interconnect wires that comprise 
each set of nets that implement the aforementioned features within our surrogate SoC.

\subsection{Effectiveness}
\label{subsection:eval_effectiveness}
We first evaluate the effectiveness of \dpar{} in thwarting the insertion of hardware Trojans at 
fabrication time. We compare the degree of protection provided by \dpar{} with that provided by 
deploying the current state-of-the-art \textit{preventive} defense suggested by Ba {\em et 
al.}~\cite{ba2015hardware,ba2016hardware}. This placement-based defense involves filling as many empty 
placement sites as possible (they show filling 95\% of all placement sites is the max feasible), 
prioritizing empty sites nearest security-critical nets. We use our automated toolchain 
(\S\ref{subsection:implementation_toolchain}) to deploy both types of guard wires (existing and designed-in). 
We assume the best case scenario for Ba {\em et al.'s} placement 
defense~\cite{ba2015hardware,ba2016hardware} by filling 95\% of the device layer with inverter cells---the 
smallest cells in our 45\,$nm$ cell library, for fine grain filling.

\label{subsection:eval_effectiveness_icas}
We use the ICAS framework~\cite{icas} to quantify the effectiveness of each defense. 
ICAS analyzes the physical layout of an IC (encoded in a GSDII file), and computes security metrics 
detailing the IC layout's fabrication-time attack surface. Namely, it computes three metrics: 1) 
\textit{trigger space}, 2) \textit{net blockage}, and 3) \textit{route distance}. 
The \textit{\textbf{trigger space}} metric characterizes the open space on the device layer (empty 
placement sites) available for an attacker to add their Trojan components. The \textit{\textbf{net 
blockage}} metric computes the percentage of surface area of identified nets that are blocked by 
other circuit components or wiring. Lastly, the \textit{\textbf{route distance}} metric computes the 
minimal distance between unblocked identified nets and unused placement sites that an adversary 
would have to route a rogue Trojan wire to ``connect'' the hardware Trojan to the host IC. 
The trigger space metric quantifies the difficulty of performing \textit{Trojan Placement}, the net 
blockage quantifies the difficulty of performing \textit{Trojan/Victim Integration}, and the route 
distance metric quantifies the difficulty of performing \textit{Intra-Trojan Routing} 
(\S\ref{subsection:fab_time_attacks}). Of the three ICAS metrics, the \textit{net blockage} metric is 
most adept to quantifying the deployability of each guard wire type (existing and designed-in), i.e., how 
effective each guard wire type is at shielding all targeted nets. Alternatively, the \textit{route 
distance} metric is the adept at comparing \dpar{} with Ba \textit{et al.}'s placement defense, as it is 
essentially a combination of the \textit{trigger space} metric---an entirely placement-focused metric---and 
the \textit{net-blockage} metric---an entirely routing-focused metric. Therefore, we utilize 
these two ICAS metrics in the following evaluation.

\begin{figure}[t]
\centering
\includegraphics[width=0.45\textwidth]{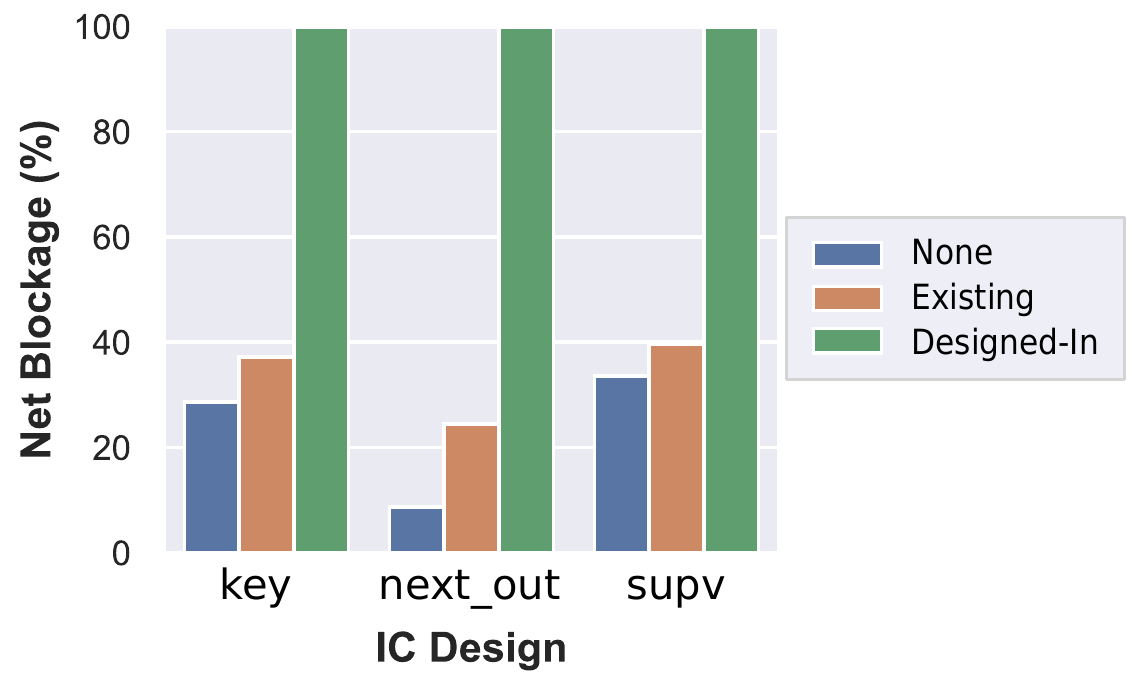}
\vspace*{-0.1in}
\caption{\footnotesize Plot of the \textit{net blockage}~\cite{icas} computed across three 
different sets of targeted nets within our SoC layout, with and without guard wires.}
\label{fig:gwire_type_nb}
\figline{}
\vspace*{-0.2in}
\end{figure}

\begin{figure*}[t]
\centering
\includegraphics[width=0.7\textwidth]{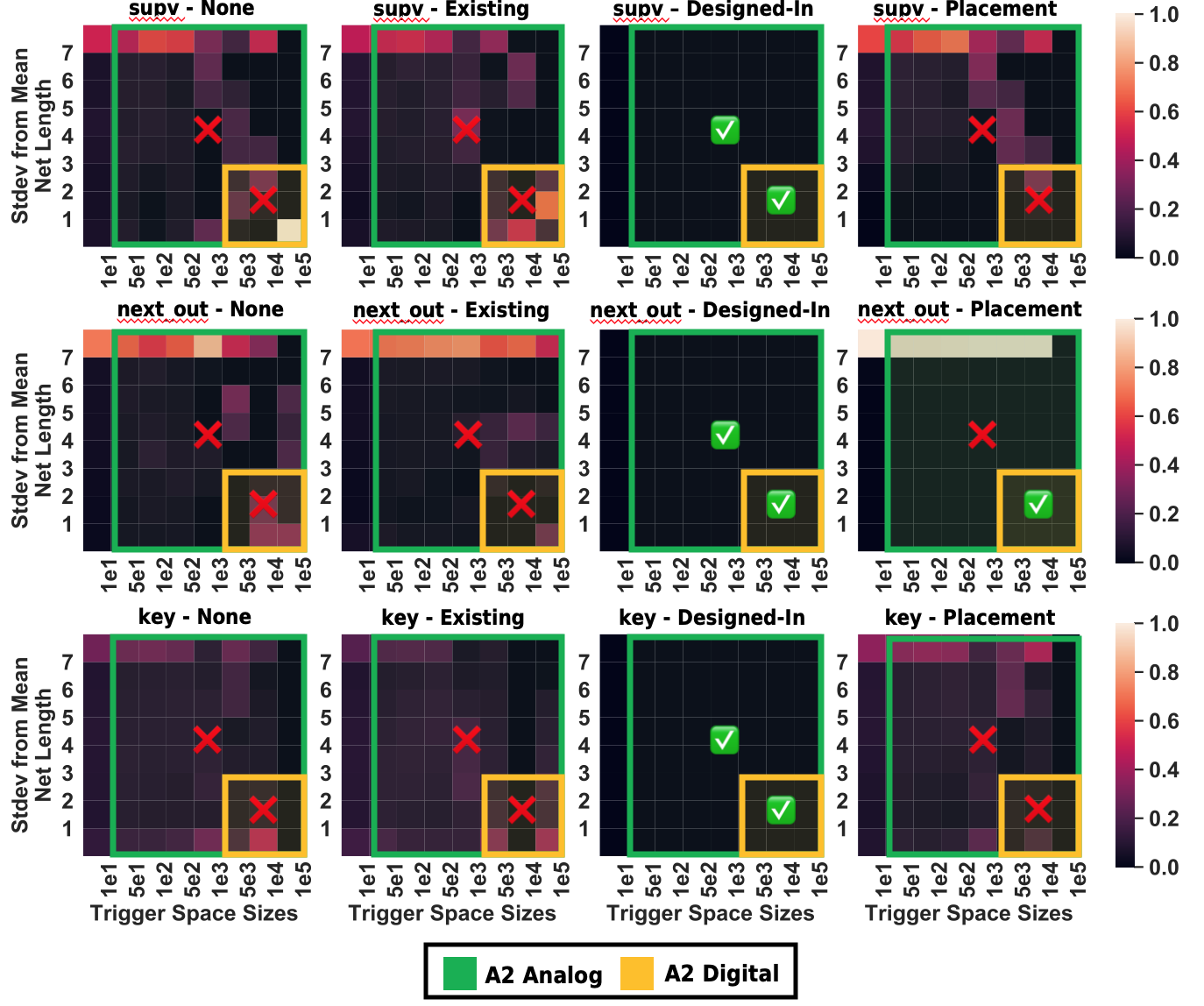}
\caption{\footnotesize Plot of the ICAS \textit{route distance} metric~\cite{icas} computed across four 
different layouts of each core within our surrogate SoC, with and without guard wires and Ba {\em et al.}'s 
defensive placement~\cite{ba2015hardware,ba2016hardware}. Each heatmap illustrates the percentage of 
(targeted net,~trigger-space) pairs (possible Trojan layout implementations) of varying distances 
apart. The heatmaps are intended to be analyzed by column, as each column encodes a histogram of possible 
attack configurations with trigger-spaces of a given size range (X-axis). Route distances (Y-axis) are 
displayed in terms of standard deviations from mean net length in each respective design. Heatmaps that are 
completely dark indicate no possible attack configurations exist, i.e., no placement/routing resources to 
insert any Trojan. Overlaid on each heatmap are rectangles indicating regions on the heatmap a given A2 
Trojan (Tab.~\ref{table:a2_trojan_designs}) may exploit, and markers (checks and x-marks) indicating if a 
non-zero number of specific Trojan layout implementations are possible.}
\label{fig:gwire_type_rd}
\figline{}
\vspace*{-0.2in}
\end{figure*}

\textbf{Net Blockage Results.}
Both existing and designed-in guard wires attempt to block targeted nets to prevent 
attackers from attaching rogue wires to them, thus minimizing/eliminating possible 
\textit{Victim/Trojan Integration} points (\S\ref{subsection:fab_time_attacks}). 
We use the \textit{net blockage} metric to compute the surface-area-coverage differences 
between existing and designed-in guard wires. Fig.~\ref{fig:gwire_type_nb} compares the net blockage 
computed across three total IC layouts of the same SoC design, including: three guard wires 
variations---without guard wires, with existing guard wires, and with designed-in guard wires---across 
three different sets of targeted nets. All net-blockage results are with respect to each set 
of targeted nets in the SoC. 

Across all three sets of targeted nets, designed-in guard wire provide more protection than 
existing guard wires, as expected. Specifically, for all nets, designed-in guard wires achieve 
100\% net blockage. This means that there is no place on any targeted net within the SoC 
where an attacker can attach a rogue wire. Existing guard wires are unable to achieve 100\% coverage 
due mainly to having to meet their own routing constraints which prevents our tool from locating 
enough nets to block all surfaces of all targeted nets, making them ineffective at thwarting attacks.

\textbf{Route Distance Results.}
\label{subsection:eval_effectiveness_rd_results}
Since \dpar{} only limits the routing resources needed to insert a Trojan at fabrication time, it is 
vital to understand how \dpar{} reduces the overall fabrication-time attack surface, i.e., both Trojan 
routing \textit{and} placement resources.
We use the \textit{route distance} metric to locate all possible combinations 
of unused placement sites and unblocked targeted nets---i.e., all possible Trojan attack 
configurations~\cite{icas}. We use the \textit{route distance} metric to illustrate the attack surface 
across each core within our SoC where that contains the root net of interest. We analyze the 
\textit{route distance} metric with respect to each containing core, as it is common practice for IC 
layout engineers to lay out each core separately, before integrating them, plus this increases the clarity of presentation.

Fig.~\ref{fig:gwire_type_rd} shows the \textit{route distance} metric as computed across all three containing 
cores, with and without layout-level defenses including: 1) \dpar{} (both existing and designed-in 
guard wires) and 2) defensive placement. Each heatmap is intended to be analyzed column-wise, where each 
column is a histogram of the distances between unblocked targeted nets and trigger-spaces\footnote{Trigger  spaces  are  contiguous  groups  of  placement  sites  that  are  empty, or contain (removable) capacitive fill cells~\cite{icas}} within 
a size range. Namely, each heatmap illustrates the fabrication-time attack surface of each IC layout. 
If a circuit has no attack configurations, i.e., all targeted nets are blocked or there are no 
trigger-spaces, the route distance heatmap is completely dark (column ratios of 0). If it is impossible to 
eradicate all attack configurations, the most secure layout for such a circuit would have maximum distances 
between unblocked targeted net and trigger-spaces, i.e., a heatmap with the top row the lightest 
color (top row ratios of 1). This is because larger distances increase the signal delay for the hardware 
Trojan; increasing the challenge of the attacker to meet timing constraints for their attack. Overlaid on 
each heatmap are rectangles indicating the region of the attack surface that is exploitable by the 
color-coded Trojan, and check- or x-marks indicating whether any possible attack configurations exist 
for that attack. A check-mark indicates there are zero possible Trojan layouts (success)), where an x-mark 
indicates the opposite (vulnerable).

Designed-in guard wires outperform existing guard wires and placement-centric defenses. For all three example attack payloads, designed-in guard wires were able to close the fabrication-time attack-surface by 
\textit{completely} blocking all targeted nets (Fig.~\ref{fig:gwire_type_nb}). Therefore, even 
the stealthiest A2 Trojan~\cite{a2} cannot be utilized to attack the features-of-interest within 
our SoC.

\subsection{Practicality}
\label{subsection:eval_deployability}

\dpar{} is effective, but is it practical? We evaluate the cost of deploying \dpar{} across three 
exemplar security-critical features within our SoC that have been subject to attack. Specifically, we analyze the power, route density, and 
performance (timing) overheads incurred by deploying both existing and designed-in guard wires. 
Measurements are taken with respect to each feature's containing core, similar to the route 
distance measurement. While it is common to analyze power, performance, and area, of an IC design, we 
instead analyze power, performance and \textit{route density}. Area measurements refer to the device-layer 
area, i.e., width and length, since the height (number of routing layers) is fixed for a given process 
technology. Since \dpar{} does not require additional logic gates, we do not increase the width and height 
(area) of the core area, rather \dpar{} alters the total wire length in the design. Thus, measuring routing
density overhead is more meaningful. We use the built-in features of Cadence tools to compute these
overheads. 

Fig.~\ref{fig:defensive_routing_stats} details our results. 
Power and timing overheads were both less than 1\%. In some cases, the timing was better for the guard 
wire designs. This is expected as \dpar{} does not require any additional logic gates, nor lengthen 
existing wires. Rather, the guard wires increase routing constraints that can push the PaR CAD tool to 
produce more optimal routing solutions. The route density overhead was less than 1\% for all existing guard 
wires, and similar for designed-in guard wires when the number of targeted nets to guard is small, 
namely the \textit{supv} and \textit{next\_out} nets. Intuitively, the more guard wires inserted, the 
higher the routing density increase. Keeping route density low is important to ensure automated CAD tools 
can route each design. However, even though all layouts targeted a placement density (density of logic 
gates on the device layer) of 60--80\%, route density was relatively low even with guard wires. 
This was due to the characteristics of the designs and process technology (i.e., back-end-of-line metal 
stack option).

It's worth noting, that in addition to low power, performance, and area overheads, deploying \dpar{} guard wires has minimal impact on the run-time of layout CAD tools. Without DR, the tools lay out each SoC core in less than 10 minutes, and with DR they lay out each core in less than 11 minutes. Tool run-time overheads are more impacted by the magnitude of features requiring protection than on circuit complexity.

\subsection{Threat Analysis of Bypass Attacks}
\label{subsection:eval_threat_analysis}

Lastly, we provide a threat analysis of \dpar{}. Recall, of the three ways an attacker can bypass 
\dpar{} guard wires to carry out a fabrication-time attack (Fig.~\ref{fig:gw_attacks} and 
\S\ref{subsection:gw_bypass_attacks}), the \textit{jog} attack is the stealthiest. An attacker mounts a 
jog attack by \textit{jogging}, or moving, a portion of a guard wire to a nearby routing track, 
in order to make room for a rogue Trojan wire to attach to a targeted net
(Fig.~\ref{fig:gw_attacks}C). In such an attack, the guard wire is \textit{lengthened}, or \textit{bends} are added/removed. To evaluate the detectability of such an attack, we ask three questions: 
\begin{enumerate}[topsep=0.5ex,itemsep=0.5ex,partopsep=1ex,parsep=0.5ex]
    \item \textit{What is the smallest jog attack, i.e., the minimum alteration in a guard wire's 
        length and/or number of bends?}
    \item \textit{Is the smallest jog attack masked by process variation?}
    \item \textit{Can modern TDR detect the smallest jog attacks?}
\end{enumerate}

\textbf{Smallest Jog Attack.}
The minimum jog attack is to jog a top 
(or bottom) guard wire to an adjacent routing track, and attach to the targeted net from above 
(or below) with a via, as illustrated in Fig.~\ref{fig:gw_attacks}C. This edit either increases the length of the guard wire, or adds/removes bends---impedance discontinuities---in the guard wire to keep its overall length unchanged. This edit is minimal because the minimal metal pitch (MMP), or (horizontal) distance between the centers of adjacent routing tracks on the \textit{same routing layer}, is much smaller than the (vertical) distance between overlapping routing tracks on \textit{adjacent routing layers}. Specifically, the smallest jog attack would either: 1) increase a guard wire's length by: $L_{attack} = 2 * MMP_{r}$, where $MMP_{r}$ is the MMP on layer $r$, as defined in the design rules of a given process technology, or 2) add/remove bend(s) in the guard wire that are at least a distance of $L_{attack}$ apart from existing bends. \textit{In either case, a feature resolution---of overall length or length between bends---of $L_{attack}$ is required to detect the smallest jog attack}. Table~\ref{table:min_gwire_edits} summarizes the \emph{minimal-attack-edits} ($L_{attack}$ distances), to a guard wire's features an attacker must make to bypass \dpar{}, according to the 45\,$nm$ process technology we target in this study. 

\begin{table}
\centering
\caption{\footnotesize Minimum guard wire jog attack (Fig.~\ref{fig:gw_attacks}C) edit--distances for each routing layer in the IBM 45\,$nm$ SOI process technology.}
\label{table:min_gwire_edits}
\begin{tabular}{c c c c c}
\multicolumn{1}{c}{\multirow{1}{*}{\footnotesize \textbf{Routing}}} & 
\multicolumn{1}{c}{\multirow{1}{*}{\footnotesize \textbf{Min Wire}}} &
\multicolumn{1}{c}{\multirow{1}{*}{\footnotesize \textbf{Min Metal}}} &
\multicolumn{1}{c}{\multirow{1}{*}{\footnotesize \textbf{Min Attack}}} &
\multicolumn{1}{c}{\multirow{1}{*}{\footnotesize \textbf{TDR}}} \\

\multicolumn{1}{c}{\multirow{1}{*}{\footnotesize \textbf{Layer}}} & 
\multicolumn{1}{c}{\multirow{1}{*}{\footnotesize \textbf{Spacing ($um$)}}} &
\multicolumn{1}{c}{\multirow{1}{*}{\footnotesize \textbf{Pitch ($um$)}}} &
\multicolumn{1}{c}{\multirow{1}{*}{\footnotesize \textbf{Edit ($um$)}}} &
\multicolumn{1}{c}{\multirow{1}{*}{\footnotesize \textbf{Detectable?}}} \\ 
\hline
\hline
\footnotesize 1  & \footnotesize 0.07 & \footnotesize 0.14 & \footnotesize 0.28 & \footnotesize \cellcolor{GreenHLight!30}\cmark \\
\footnotesize 2  & \footnotesize 0.07 & \footnotesize 0.14 & \footnotesize 0.28 & \footnotesize \cellcolor{GreenHLight!30}\cmark \\
\footnotesize 3  & \footnotesize 0.07 & \footnotesize 0.14 & \footnotesize 0.28 & \footnotesize \cellcolor{GreenHLight!30}\cmark \\ \hline
\footnotesize 4  & \footnotesize 0.09 & \footnotesize 0.19 & \footnotesize 0.38 & \footnotesize \cellcolor{GreenHLight!30}\cmark \\
\footnotesize 5  & \footnotesize 0.09 & \footnotesize 0.19 & \footnotesize 0.38 & \footnotesize \cellcolor{GreenHLight!30}\cmark \\ \hline
\footnotesize 6  & \footnotesize 0.14 & \footnotesize 0.28 & \footnotesize 0.56 & \footnotesize \cellcolor{GreenHLight!30}\cmark \\
\footnotesize 7  & \footnotesize 0.14 & \footnotesize 0.28 & \footnotesize 0.56 & \footnotesize \cellcolor{GreenHLight!30}\cmark \\ \hline
\footnotesize 8  & \footnotesize 0.80 & \footnotesize 1.60 & \footnotesize 3.20 & \footnotesize \cellcolor{GreenHLight!30}\cmark \\
\footnotesize 9  & \footnotesize 0.80 & \footnotesize 1.60 & \footnotesize 3.20 & \footnotesize \cellcolor{GreenHLight!30}\cmark \\ \hline
\footnotesize 10 & \footnotesize 2.00 & \footnotesize 4.00 & \footnotesize 8.00 & \footnotesize \cellcolor{GreenHLight!30}\cmark
\end{tabular}
\vspace*{-0.2in}
\end{table}

\textbf{Process Variation vs. Smallest Jog Attack.}
Assume for a moment that we can measure the of overall length, or length between bends, of a guard wire to infinite accuracy. Even then, detecting the smallest jog attack requires the minimal attack edit distance, $L_{attack}$, be discernable from deviations between simulated and fabricated guard wire lengths due to process variation. Fortunately, \textbf{$L_{attack}$ is larger than the \underline{worst-case} manufacturing process variation} in a guard wire's length. Namely, with $L_{design}$ as the designed length of the guard wire, and $L_{wc\_error}$, as the worst-case manufacturing error in the actual guard wire's length (\texttt{+} or \texttt{-}): 

\vspace*{-0.2in}
\begin{equation}\label{eq:process_variation}
    \begin{aligned}
        L_{design} - L_{wc\_error} + L_{attack} &> L_{design} + L_{wc\_error}
    \end{aligned}
\end{equation}

For a guard wire on routing layer $r$, the \textit{worst-case} manufacturing error, $L_{wc\_error}$, can be deduced from the manufacturing design rules as: 

\vspace*{-0.2in}
\begin{equation}\label{eq:wc_process_variation}
    \begin{aligned}
        L_{wc\_error} &= 2 * \frac{min\_spacing_{r}}{2} = min\_spacing_{r}
    \end{aligned}
\end{equation}
where $min\_spacing_{r}$ is the minimum required spacing surrounding a wire routed on metal layer, $r$.

\begin{figure}[t]
\centering
\vspace*{-0.1in}
\includegraphics[width=0.37\textwidth]{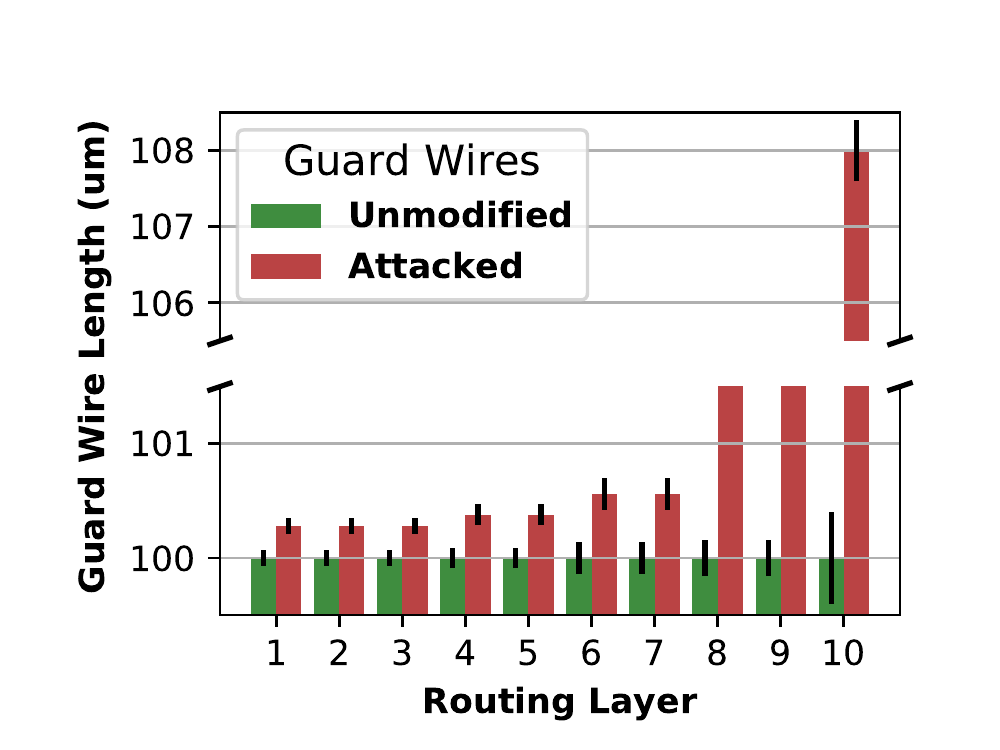}
\vspace*{-0.2in}
\caption{\footnotesize Worst-case manufacturing process variation (error bars) effect on unmodified 
and minimal jog attacks on 100-micron guard-wires.}
\label{fig:pv_simulation}
\figline{}
\vspace*{-0.2in}
\end{figure}

We illustrate this in Fig.~\ref{fig:pv_simulation}, where we plot the minimum length differences between 
unmodified (un-attacked) and minimally-jogged (attacked) guard wires, overlaid with error bars indicating 
the worst-case range of variation in a guard wires fabricated length caused by process variation. 
Even in the worst case, across all routing layers, unmodified vs attacked guard wires are discernible.

\textbf{Attack Detection with TDR.}
When IC interconnects are injected with a pulsed waveform with a rise time less than twice the 
propagation delay of the interconnect, they behave like transmission lines 
(Eq.~(\ref{eq:transmission_line_model})). Hence, time-domain reflectometry (TDR) can be used to 
measure several characteristics of designed-in guard wires to ensure they have not been tampered with 
(\S\ref{subsection:background_tdr}). Specifically, 
the \textit{lengths} of each guard wire, or \textit{lengths between bends} on each guard wire, are computed by measuring the reflection time(s) of a single incident rising pulse applied to the guard wires under test. Once measured, the lengths can be compared with that predicted by a 3D electromagnetic field solver~\cite{lim2014characterization} to detect if they have been altered. While modeling \textit{all} interconnects within a large complex IC using a field solver is computationally impractical, it is \emph{practical} to analyze only a small subset of interconnects, e.g., the guard wires and surrounding circuit structures~\cite{nagel2011contact}.

Prior work demonstrates terahertz TDR systems
\cite{cai2010electro,tay2012advanced,nagel2011contact,teraview} capable of measuring the propagation 
delay of an interconnect to a resolution of $\pm$2.6\,femptoseconds ($fs$). Such systems utilize laser-driven
optoelectronic measurement techniques to achieve such high resolutions. 
According to the ideal transmission line model~\cite{sutherland1999edge}, the propagation delay, 
$T_{pd}$, is a function of the dielectric constant, $D_k$, speed of light, $C$, and 
\textbf{length of the transmission line (guard wire)}, $L_{gw}$, as shown in Eq.~(\ref{eq:prop_delay}).

\vspace*{-0.15in}
\begin{equation}
\label{eq:prop_delay}
    T_{pd} = L_{gw} * \frac{\sqrt{D_k}}{C}
\end{equation}
\vspace*{-0.15in}

TDR is the ideal tamper detection tool as process variation has no impact on its 
\textit{\underline{accuracy}}. Knowing the dielectric constant, $D_k$, of the 
insulating material surrounding the guard wires---the inter-layer dielectric (ILD)---is \textit{all} that is required to compute their lengths, or the lengths between their bends (Eq.~(\ref{eq:prop_delay})). Since, the dielectric constant of the ILD is \textbf{not} dependent on its geometric properties, it is well controlled~\cite{boning2000models}. 


\begin{figure}[t]
\centering
\includegraphics[width=0.35\textwidth]{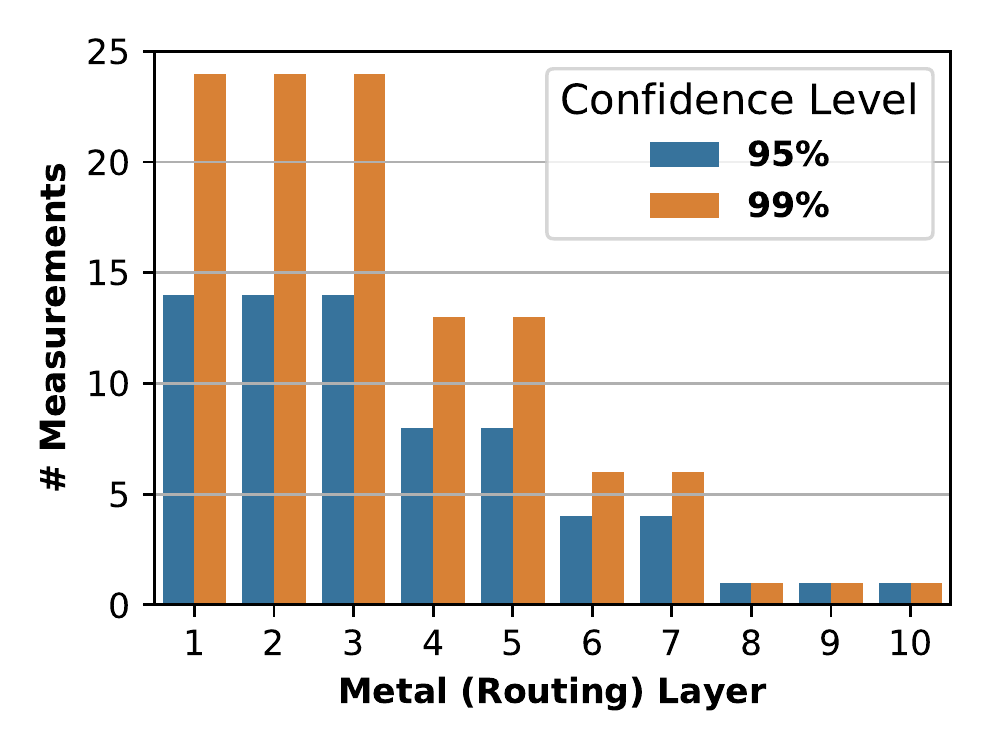}
\vspace*{-0.20in}
\caption{\footnotesize Number of TDR measurements required to detect the smallest jog attacks (Table~\ref{table:min_gwire_edits}) with 95\% and 99\% confidence, per layer.}
\label{fig:tdr_simulation}
\figline{}
\vspace*{-0.2in}
\end{figure}

Using the TDR propagation delay model described in Eq.~(\ref{eq:prop_delay}), and the previously 
studied resolution of optoelectrical terahertz 
TDR~\cite{cai2010electro,tay2012advanced,nagel2011contact,teraview}, we simulate the detection of the smallest jog attacks on guard wires across every routing layer in our target 45\,$nm$ process. Namely, we simulate the difference in reflection times observed for single pulse TDR waveforms applied to (unmodified) guard wires that are 100 microns long, compared to the reflection time observed from similar guard wires that have been lengthened by the minimal attack edit distances, $L_{attack}$, across each routing layer (Table~\ref{table:min_gwire_edits}). We assume a 
dielectric constant of 3.9, the nominal dielectric constant of silicon dioxide~\cite{kingon2000alternative}. 
Taking into account a (Gaussian) standard error (across reflection time measurements) of $\pm\,2.6$ $fs$, 
as reported by~\cite{nagel2011contact}, we compute the minimum number of TDR measurements required to 
discriminate an unmodified guard wire from an attacked guard wire with confidence levels of 95\% and 99\%. 
We plot these results in Figure~\ref{fig:tdr_simulation}. Our results demonstrate that existing terahertz 
TDR systems are capable of detecting the smallest jog attacks across all routing layers 
(Table~\ref{table:min_gwire_edits}) in our target 45\,$nm$ process, requiring at most 14 and 24 TDR 
measurements to achieve confidence levels of 95\% and 99\%, respectively.

\section{Discussion}\label{section:discussion}
\dpar{} aims to prevent fabrication-time Trojan attacks that target specific security-critical features in an IC design. Experiments on real circuit layouts of a SoC containing show that \dpar{} is effective, deployable, and tamper-evident. Discussed below are the limitations, scalability, signal integrity impact, flexibility, and extensibility of \dpar{}.

\textbf{Limitations.} 
\dpar{} is a mitigation strategy for hardware designs where only a subset of the design is 
security-critical~\cite{glift,specs15}. As our evaluation results show, 
the deployability and performance overhead 
of \dpar{} is low when the overall security-critical wire length is low. 
If \textit{every} wire in a design is security-critical, then \dpar{} is not a good defensive strategy;
in fact, the motive for outsourcing fabrication in such scenarios is tenuous.
If fabrication must be outsourced, we recommend alternative
mitigation strategies such as those proposed in
\cite{ba2015hardware,ba2016hardware,xiao2013bisa,linscott2018swan,imeson2013split}. 
The tradeoff is that these strategies have limited deployability, and a large, fixed, 
performance overhead that make them impractical for designs that require only 
a subset of security-critical functionality be protected.


\textbf{Scalability.}
There are two notions of scalability to address. The first is scalability with regard to \textit{routability}. Routing guard wires alongside security-critical wires can impact the routability of a layout, if the \textbf{\textit{percentage of overall wire length}} to guard is large. By placing and routing security-critical components and wires first, before any other portions of the circuit (\S\ref{subsection:implementation_par}), we are able to minimize security-critical wire length. This makes security-critical wire length scale with the total length of security-critical wires, as opposed to the larger design. As we see when going from OR1200 and RISC-V class processor to modern x86-64 processors, the proportion of security-critical functionality (hence wires) decreases as relatively more transistors are spent on performance. Furthermore, even when the security-critical wire length percentage is large, as is the case within the AES core (Fig.~\ref{fig:defensive_routing_stats}-Route Density), we are able to guard over 1000 nets with little impact on power or performance. In fact, this is one reason we select AES as a benchmark (even though it is arguably entirely security-critical): its key-bit nets exhibit a unique quality that stress tests \dpar{}. Specifically, they are global, highly-connected routes, i.e., they are orders of magnitude longer than any other nets in the layout and most of the design uses them.

The second notion of scalability is with regard to the detection of bypass attacks. Although Moore's law is near its limit, transistors continue to shrink. Only three companies in the world are capable of manufacturing 7--10\,$nm$ transistors~\cite{gf_7nm}. It is, therefore, vital for \dpar{} to scale with process technology. With respect to deletion attacks (Fig.~\ref{fig:gw_attacks}A), \dpar{} scales with process technology advances as measuring interconnect continuity does not differ across process technologies. With respect to move attacks (Fig.~\ref{fig:gw_attacks}B), \dpar{} scales with process technology advances as cross-talk is amplified when interconnects are smaller and more densely packed. Lastly, with respect to jog attacks, \dpar{} also scales, as TDR capabilities directly scale with microelectronic feature sizes, i.e., faster transistors translates to faster TDR rise times.

\textbf{Signal Integrity Impact.}
Routing long wires parallel to targeted nets increases coupling capacitance, thus creating cross-talk between the guard wires and the targeted nets they protect. However, designed-in guard wires are not actively driven (and can be grounded) during normal chip operation. Thus, cross-talk is not an issue---in fact, designed-in guard wires decrease cross-talk by acting as shields between targeted nets and the rest of the circuit. 

\textbf{Defense-in-Depth.}
While \dpar{} alone can thwart even the stealthiest fabrication-time attacks, its low deployment costs also enable defense-in-depth. Layering \dpar{} with other preventive measures, such as Ba {\em et al.}'s defensive placement~\cite{ba2015hardware,ba2016hardware}, provides an additional layer of protection.

\textbf{Extensibility of CAD Tools.} 
Our \dpar{} deployment framework (\S\ref{section:implementation}) is built on top of a commercial IC CAD tool~\cite{innovus} and an open-source VLSI analysis tool~\cite{icas}. Extending \dpar{} to work across other commercial IC layout CAD tools involves incorporating support for each vendor's CAD tool APIs. We foresee \dpar{} deployed as an integrated component of commercial VLSI CAD tools as they focus more on IC security.

\section{Related Work}\label{section:related_work}
Fabrication-time attacks and defenses have been extensively studied. Attacks have spanned the trade-space of footprint size, stealth, and controllability. Specifically, some attacks have demonstrated stealth and controllability, at the cost of large footprints~\cite{lin2009trojan,becker2013stealthy,king08}, while others have demonstrated small (or non-existent) footprints, at the cost of controllability and stealth~\cite{shiyanovskii2010process,kumar2014parametric}. The most formidable attack---the A2 attack~\cite{a2}---has demonstrated all three: small footprint, stealth, and controllability.
We highlight a few notable attacks and defenses below.

On the defensive side, there are two main strategies: detective or preventive. Most prior work has 
focused on detective strategies, while few works have focused on preventive strategies. Detective strategies 
involve side-channel analysis~\cite{agrawal2007trojan,jin2008hardware,balasch2015electromagnetic,narasimhan2011tesr}, imaging~\cite{zhou2015detecting,adato2016rapid}, and on-chip 
sensors~\cite{li2008speed,forte2013temperature,hou2018r2d2}. Until \dpar{}, preventive measures have been 
placement-focused \cite{xiao2013bisa,ba2015hardware,ba2016hardware}.

\textbf{Fabrication-time Attacks.}\label{section:rr_attacks}
The first fabrication-time insertion of a hardware Trojan was developed by Lin 
{\em et al.}~\cite{lin2009trojan} who proposed a Trojan designed to leak information over a deliberately 
created side channel. Specifically, they designed and implemented a hardware Trojan, with a footprint of 
approximately 100 logic gates, to create an artificial power side channel for leaking cryptographic keys. 
Albeit unique at the time, today such a large footprint makes the attack detectable via side channel 
defenses~\cite{agrawal2007trojan,balasch2015electromagnetic,forte2013temperature}.

The most lethal fabrication-time attack is the A2 Trojan, developed by Yang {\em et al.}~\cite{a2}. 
The A2 Trojan utilizes analog components to build a counter-based trigger circuit with a footprint of 
less than the size of one flip-flop. Its complex triggering mechanism makes it stealthy, i.e., unlikely to 
accidentally deploy during post-fabrication functional testing or under normal chip operation, yet is 
controllable from user-level software. Its unique design makes it the only Trojan to evade all detection 
schemes, except \dpar{}. 

\textbf{Fabrication-time Defenses.}\label{section:rr_defenses}
The first side-channel detection scheme was proposed by Agrawal {\em et al.}~\cite{agrawal2007trojan}. 
They used power, temperature, and electromagnetic (EM) side-channel measurements to record a fingerprint 
of a ``golden'' IC during normal, and compared this fingerprint to one acquired from an untrusted IC. 
Similarly, Jin {\em et al.}~\cite{jin2008hardware} create a timing-based fingerprint obtained by measuring 
the output delays resulting from applying various input combinations to a given IC. While side-channel 
detection schemes are effective against hardware Trojans with large footprints, they fail at detecting 
Trojans like A2~\cite{a2}, whose side-channel signatures are well below operational noise margins.


Of all fabrication-time Trojan defenses, R2D2~\cite{hou2018r2d2} is the only one that claims to detect the A2 Trojan. R2D2 works by using on-chip sensors to monitor the toggling frequency of a select few security-critical signals within the design. If the toggling rate of any security-critical signals exceed a pre-determined threshold, then an alarm signal is activated to indicate an A2 Trojan may have been triggered. The crux of this approach is that, unlike \dpar{} guard wires, the hardware used to construct the toggle frequency monitors \textit{is not tamper-evident}. There is no way to tell if a foundry-side attacker disabled the R2D2 hardware while inserting her Trojan.

\section{Conclusion}\label{section:conclusion}
\dpar{} is a routing-centric preventive defense against additive fabrication-time Trojans that target 
security-critical hardware features. It makes routing Trojan wires to, or directly adjacent to, 
attacker-targeted wires in a victim IC intractable by surrounding their surfaces with tamper-evident guard 
wires. We propose the use of designed-in guard wires in conjunction with post-fabrication terahertz time-domain reflectometry (TDR) analysis to detect \textit{all} bypass attacks we contrive (\textit{deletion}, \textit{move}, and \textit{jog} attacks). We develop an automated toolchain for deploying \dpar{} guard wire. 
Lastly, we evaluate the effectiveness, deployability, and tamper-evidence of \dpar{} at securing 
multiple security-critical features within an SoC that have been subject to attack by existing hardware Trojans.
Our results show that \dpar{} thwarts the insertion of even the stealthiest known 
additive hardware Trojan---the A2 Trojan---with power, timing, and area overheads of $\approx$\,1\%.

\section*{Acknowledgment}
We thank Brian Tyrrell, Matt Guyton, and other members of the MIT Lincoln Laboratory community for their thoughtful feedback that enhanced the quality of our work.

DISTRIBUTION STATEMENT A. Approved for public release. Distribution is unlimited. This material is based upon work supported under Air Force Contract No. FA8702-15-D-0001. Any opinions, findings, conclusions or recommendations expressed in this material are those of the author(s) and do not necessarily reflect the views of the U.S. Air Force. © 2020 Massachusetts Institute of Technology. Delivered to the U.S. Government with Unlimited Rights, as defined in DFARS Part 252.227-7013 or 7014 (Feb 2014). Notwithstanding any copyright notice, U.S. Government rights in this work are defined by DFARS 252.227-7013 or DFARS 252.227-7014 as detailed above. Use of this work other than as specifically authorized by the U.S. Government may violate any copyrights that exist in this work.

This material is based upon work supported by the National Science Foundation Graduate Research Fellowship Program under Grant No. DGE 1256260. Any opinions, findings, and conclusions or recommendations expressed in this material are those of the author(s) and do not necessarily reflect the views of the National Science Foundation.

\bibliographystyle{plain}
{\footnotesize
\bibliography{sections/bibliography.bib}}

\begin{thebibliography}{10}

\bibitem{adato2016rapid}
Ronen Adato, Aydan Uyar, Mahmoud Zangeneh, Boyou Zhou, Ajay Joshi, Bennett
  Goldberg, and M~Selim Unlu.
\newblock Rapid mapping of digital integrated circuit logic gates via
  multi-spectral backside imaging.
\newblock {\em arXiv:1605.09306}, 2016.

\bibitem{agrawal2007trojan}
Dakshi Agrawal, Selcuk Baktir, Deniz Karakoyunlu, Pankaj Rohatgi, and Berk
  Sunar.
\newblock Trojan detection using {IC} fingerprinting.
\newblock In {\em IEEE Symposium on Security and Privacy (S\&P)}, 2007.

\bibitem{alkabani2008designer}
Yousra Alkabani and Farinaz Koushanfar.
\newblock Designer’s hardware trojan horse.
\newblock In {\em IEEE International Workshop on Hardware-Oriented Security and
  Trust (HOST)}, 2008.

\bibitem{ba2016hardware}
Papa-Sidy Ba, Sophie Dupuis, Manikandan Palanichamy, Giorgio Di~Natale, Bruno
  Rouzeyre, et~al.
\newblock Hardware trust through layout filling: a hardware trojan prevention
  technique.
\newblock In {\em IEEE Computer Society Annual Symposium on VLSI (ISVLSI)},
  2016.

\bibitem{ba2015hardware}
Papa-Sidy Ba, Manikandan Palanichamy, Sophie Dupuis, Marie-Lise Flottes,
  Giorgio Di~Natale, and Bruno Rouzeyre.
\newblock Hardware trojan prevention using layout-level design approach.
\newblock In {\em European Conference on Circuit Theory and Design (ECCTD)},
  2015.

\bibitem{bakoglu1990circuits}
Halil~B Bakoglu.
\newblock Circuits, interconnections, and packaging for vlsi., 1990.

\bibitem{balasch2015electromagnetic}
Josep Balasch, Benedikt Gierlichs, and Ingrid Verbauwhede.
\newblock Electromagnetic circuit fingerprints for hardware trojan detection.
\newblock In {\em IEEE International Symposium on Electromagnetic Compatibility
  (EMC)}, 2015.

\bibitem{beaumont2011hardware}
Mark Beaumont, Bradley Hopkins, and Tristan Newby.
\newblock Hardware trojans-prevention, detection, countermeasures (a literature
  review).
\newblock Technical report, Defence Science and Technology Organization
  Edinburgh (Australia), 2011.

\bibitem{becker2013stealthy}
Georg~T Becker, Francesco Regazzoni, Christof Paar, and Wayne~P Burleson.
\newblock Stealthy dopant-level hardware trojans.
\newblock In {\em International Workshop on Cryptographic Hardware and Embedded
  Systems (CHES)}, 2013.

\bibitem{boning2000models}
Duane Boning and Sani Nassif.
\newblock Models of process variations in device and interconnect.
\newblock {\em Design of high performance microprocessor circuits}, 2000.

\bibitem{innovus}
{Cadence Design Systems}.
\newblock Innovus implementation system.
\newblock
  \url{https://www.cadence.com/content/cadence-www/global/en\_US/home.html}.

\bibitem{cai2010electro}
Yongming Cai, Zhiyong Wang, Rajen Dias, and Deepak Goyal.
\newblock Electro optical terahertz pulse reflectometry—an innovative fault
  isolation tool.
\newblock In {\em Electronic Components and Technology Conference (ECTC), 2010
  Proceedings 60th}, 2010.

\bibitem{chakraborty2009hardware}
Rajat~Subhra Chakraborty, Seetharam Narasimhan, and Swarup Bhunia.
\newblock Hardware trojan: Threats and emerging solutions.
\newblock In {\em IEEE International High Level Design Validation and Test
  Workshop (HLDVT)}. IEEE, 2009.

\bibitem{chen2006nondestructive}
Ming-Kun Chen, Cheng-Chi Tai, and Yu-Jung Huang.
\newblock Nondestructive analysis of interconnection in two-die bga using tdr.
\newblock {\em IEEE Transactions on Instrumentation and Measurement}, 2006.

\bibitem{forte2013temperature}
Domenic Forte, Chongxi Bao, and Ankur Srivastava.
\newblock Temperature tracking: An innovative run-time approach for hardware
  trojan detection.
\newblock In {\em IEEE/ACM International Conference on Computer-Aided Design
  (ICCAD)}, 2013.

\bibitem{hayden1994characterization}
Leonard~A Hayden and Vijai~K Tripathi.
\newblock Characterization and modeling of multiple line interconnections from
  time domain measurements.
\newblock {\em IEEE Transactions on Microwave Theory and Techniques}, 1994.

\bibitem{hicks10}
Matthew Hicks, Murph Finnicum, Samuel~T. King, Milo M.~K. Martin, and
  Jonathan~M. Smith.
\newblock Overcoming an untrusted computing base: Detecting and removing
  malicious hardware automatically.
\newblock In {\em IEEE Symposium on Security and Privacy (S\&P)}, 2010.

\bibitem{specs15}
Matthew Hicks, Cynthia Sturton, Samuel~T. King, and Jonathan~M. Smith.
\newblock Specs: A lightweight runtime mechanism for protecting software from
  security-critical processor bugs.
\newblock In {\em International Conference on Architectural Support for
  Programming Languages and Operating Systems}, ASPLOS, 2015.

\bibitem{hollis2006rasp}
Simon Hollis and Simon~W Moore.
\newblock Rasp: an area-efficient, on-chip network.
\newblock In {\em 2006 International Conference on Computer Design}, pages
  63--69. IEEE, 2006.

\bibitem{hollis2009pulse}
Simon~J Hollis.
\newblock Pulse generation for on-chip data transmission.
\newblock In {\em 2009 12th Euromicro Conference on Digital System Design,
  Architectures, Methods and Tools}, pages 303--310. IEEE, 2009.

\bibitem{hou2018r2d2}
Yumin Hou, Hu~He, Kaveh Shamsi, Yier Jin, Dong Wu, and Huaqiang Wu.
\newblock {R2D2}: Runtime reassurance and detection of {A2} trojan.
\newblock In {\em International Symposium on Hardware Oriented Security and
  Trust (HOST)}. IEEE, 2018.

\bibitem{hsue1997reconstruction}
Ching-Wen Hsue and Te-Wen Pan.
\newblock Reconstruction of nonuniform transmission lines from time-domain
  reflectometry.
\newblock {\em IEEE Transactions on Microwave Theory and Techniques}, 1997.

\bibitem{imeson2013split}
Frank Imeson, Ariq Emtenan, Siddharth Garg, and Mahesh Tripunitara.
\newblock Securing computer hardware using {3D} integrated circuit ({IC})
  technology and split manufacturing for obfuscation.
\newblock In {\em {USENIX} Security Symposium}, 2013.

\bibitem{jin2010dftt}
Yier Jin, Nathan Kupp, and Yiorgos Makris.
\newblock Dftt: Design for trojan test.
\newblock In {\em IEEE International Conference on Electronics, Circuits, and
  Systems (ICECS)}, 2010.

\bibitem{jin2008hardware}
Yier Jin and Yiorgos Makris.
\newblock Hardware trojan detection using path delay fingerprint.
\newblock In {\em IEEE International Workshop on Hardware-Oriented Security and
  Trust (HOST)}, 2008.

\bibitem{kelly2015detecting}
Shane Kelly, Xuehui Zhang, Mohammed Tehranipoor, and Andrew Ferraiuolo.
\newblock Detecting hardware trojans using on-chip sensors in an asic design.
\newblock {\em Journal of Electronic Testing}, 31(1):11--26, 2015.

\bibitem{king08}
Samuel~T. King, Joseph Tucek, Anthony Cozzie, Chris Grier, Weihang Jiang, and
  Yuanyuan Zhou.
\newblock Designing and implementing malicious hardware.
\newblock In {\em Proceedings of the Usenix Workshop on Large-Scale Exploits
  and Emergent Threats (LEET)}, 2008.

\bibitem{kingon2000alternative}
Angus~I Kingon, Jon-Paul Maria, and SK~Streiffer.
\newblock Alternative dielectrics to silicon dioxide for memory and logic
  devices.
\newblock {\em Nature}, 2000.

\bibitem{kumar2014parametric}
Raghavan Kumar, Philipp Jovanovic, Wayne Burleson, and Ilia Polian.
\newblock Parametric trojans for fault-injection attacks on cryptographic
  hardware.
\newblock In {\em Workshop on Fault Diagnosis and Tolerance in Cryptography
  (FDTC)}, 2014.

\bibitem{battling_fab_cycle_time}
Mark Lapedus.
\newblock Battling fab cycle times, February 2017.
\newblock \url{https://semiengineering.com/battling-fab-cycle-times/}.

\bibitem{cost_of_fab_3nm}
Mark Lapedus.
\newblock Big trouble at 3nm, June 2018.
\newblock \url{https://semiengineering.com/big-trouble-at-3nm/}.

\bibitem{gf_7nm}
Mark Lapedus.
\newblock {GF} puts 7nm on hold, August 2018.
\newblock \url{https://semiengineering.com/gf-puts-7nm-on-hold/}.

\bibitem{li2008speed}
Jie Li and John Lach.
\newblock At-speed delay characterization for ic authentication and trojan
  horse detection.
\newblock In {\em IEEE International Workshop on Hardware-Oriented Security and
  Trust (HOST)}, 2008.

\bibitem{lim2014characterization}
Jun~Jun Lim, Nor~Adila Johari, Subhash~C Rustagi, and Narain~D Arora.
\newblock Characterization of interconnect process variation in cmos using
  electrical measurements and field solver.
\newblock {\em IEEE Transactions on Electron Devices}, 2014.

\bibitem{lin2009trojan}
Lang Lin, Markus Kasper, Tim G{\"u}neysu, Christof Paar, and Wayne Burleson.
\newblock Trojan side-channels: Lightweight hardware trojans through
  side-channel engineering.
\newblock In {\em International Workshop on Cryptographic Hardware and Embedded
  Systems (CHES)}, 2009.

\bibitem{linscott2018swan}
Timothy Linscott, Pete Ehrett, Valeria Bertacco, and Todd Austin.
\newblock Swan: mitigating hardware trojans with design ambiguity.
\newblock In {\em IEEE/ACM International Conference on Computer-Aided Design
  (ICCAD)}. IEEE, 2018.

\bibitem{cep}
{MIT Lincoln Laboratory}.
\newblock Common evaluation platform.
\newblock \url{https://github.com/mit-ll/CEP}.

\bibitem{cep_v1}
{MIT Lincoln Laboratory}.
\newblock Common evaluation platform.
\newblock
  \url{https://github.com/mit-ll/CEP/tree/d19a5de3dc32d58b535f52fc9aa2cd70f95107e1}.

\bibitem{nagel2011contact}
Michael Nagel, Alexander Michalski, and Heinrich Kurz.
\newblock Contact-free fault location and imaging with on-chip terahertz
  time-domain reflectometry.
\newblock {\em Optics Express}, 2011.

\bibitem{narasimhan2011tesr}
Seetharam Narasimhan, Xinmu Wang, Dongdong Du, Rajat~Subhra Chakraborty, and
  Swarup Bhunia.
\newblock Tesr: A robust temporal self-referencing approach for hardware trojan
  detection.
\newblock In {\em IEEE International Symposium on Hardware-Oriented Security
  and Trust (HOST)}, 2011.

\bibitem{odegard1999comparative}
C~Odegard and C~Lambert.
\newblock Comparative tdr analysis as a packaging fa tool.
\newblock In {\em ISTFA 1999: 25 th International Symposium for Testing and
  Failure Analysis}, 1999.

\bibitem{or1200}
OpenCores.org.
\newblock Openrisc or1200 processor.
\newblock \url{https://github.com/openrisc/or1200}.

\bibitem{philen1982single}
Dan~L Philen, Ian~A White, Jane~F Kuhl, and Stephen~C Mettler.
\newblock Single-mode fiber otdr: Experiment and theory.
\newblock {\em IEEE Transactions on Microwave Theory and Techniques}, 1982.

\bibitem{potkonjak2009hardware}
Miodrag Potkonjak, Ani Nahapetian, Michael Nelson, and Tammara Massey.
\newblock Hardware trojan horse detection using gate-level characterization.
\newblock In {\em Proceedings of ACM/IEEE Design Automation Conference (DAC)},
  2009.

\bibitem{rostami2013hardware}
Masoud Rostami, Farinaz Koushanfar, Jeyavijayan Rajendran, and Ramesh Karri.
\newblock Hardware security: Threat models and metrics.
\newblock In {\em Proceedings of the International Conference on Computer-Aided
  Design (ICCD)}, 2013.

\bibitem{shiyanovskii2010process}
Yuriy Shiyanovskii, F~Wolff, Aravind Rajendran, C~Papachristou, D~Weyer, and
  W~Clay.
\newblock Process reliability based trojans through nbti and hci effects.
\newblock In {\em NASA/ESA Conference on Adaptive Hardware and Systems (AHS)},
  2010.

\bibitem{smolyansky2004electronic}
D~Smolyansky.
\newblock Electronic package fault isolation using tdr.
\newblock {\em ASM International}, 2004.

\bibitem{somlo1969microwave}
PI~Somlo and DL~Hollway.
\newblock Microwave locating reflectometer.
\newblock {\em Electronics Letters}, 1969.

\bibitem{dr_complexity_rising}
Ed~Sperling.
\newblock Design rule complexity rising, April 2018.
\newblock \url{https://semiengineering.com/design-rule-complexity-rising/}.

\bibitem{sugawara2014reversing}
Takeshi Sugawara, Daisuke Suzuki, Ryoichi Fujii, Shigeaki Tawa, Ryohei Hori,
  Mitsuru Shiozaki, and Takeshi Fujino.
\newblock Reversing stealthy dopant-level circuits.
\newblock In {\em International Workshop on Cryptographic Hardware and Embedded
  Systems (CHES)}, 2014.

\bibitem{sutherland1999edge}
James Sutherland.
\newblock As edge speeds increase, wires become transmission lines.
\newblock {\em EDN}, 1999.

\bibitem{tay2012advanced}
MY~Tay, L~Cao, M~Venkata, L~Tran, W~Donna, W~Qiu, J~Alton, PF~Taday, and M~Lin.
\newblock Advanced fault isolation technique using electro-optical terahertz
  pulse reflectometry.
\newblock In {\em Physical and Failure Analysis of Integrated Circuits (IPFA),
  2012 19th IEEE International Symposium on the}, 2012.

\bibitem{tehranipoor2010survey}
Mohammad Tehranipoor and Farinaz Koushanfar.
\newblock A survey of hardware trojan taxonomy and detection.
\newblock {\em IEEE Design \& Test of Computers}, 27(1), 2010.

\bibitem{teraview}
{}TeraView.
\newblock {\em Electro Optical Terahertz Pulse Reflectometry: The world’s
  fastest and most accurate fault isolation system.}

\bibitem{glift}
Mohit Tiwari, Hassan~M.G. Wassel, Bita Mazloom, Shashidhar Mysore, Frederic~T.
  Chong, and Timothy Sherwood.
\newblock Complete information flow tracking from the gates up.
\newblock In {\em International Conference on Architectural Support for
  Programming Languages and Operating Systems}, ASPLOS, pages 109--120, 2009.

\bibitem{icas}
Timothy Trippel, Kang~G. Shin, Kevin~B. Bush, and Matthew Hicks.
\newblock {ICAS}: an extensible framework for estimating the susceptibility of
  ic layouts to additive trojans.
\newblock In {\em IEEE Symposium on Security and Privacy (S\&P)}, 2020.

\bibitem{tsmc_turnaround_time}
{TSMC}.
\newblock Tsmc fabrication schedule — 2019, April 2019.
\newblock \url{https://www.mosis.com/db/pubf/fsched?ORG=TSMC}.

\bibitem{busybox}
Denys Vlasenko.
\newblock Busybox.
\newblock \url{https://www.busybox.net/}.

\bibitem{waksman2013fanci}
Adam Waksman, Matthew Suozzo, and Simha Sethumadhavan.
\newblock Fanci: identification of stealthy malicious logic using boolean
  functional analysis.
\newblock In {\em Proceedings of the ACM SIGSAC Conference on Computer \&
  Communications Security (CCS)}, 2013.

\bibitem{icarus}
Stephen Williams.
\newblock Icarus verilog.
\newblock \url{http://iverilog.icarus.com/}.

\bibitem{wolff2008towards}
Francis Wolff, Chris Papachristou, Swarup Bhunia, and Rajat~S Chakraborty.
\newblock Towards trojan-free trusted ics: Problem analysis and detection
  scheme.
\newblock In {\em Proceedings of the ACM Conference on Design, Automation and
  Test in Europe (DATE)}, 2008.

\bibitem{xiao2013bisa}
Kan Xiao and Mohammed Tehranipoor.
\newblock Bisa: Built-in self-authentication for preventing hardware trojan
  insertion.
\newblock In {\em IEEE International Symposium on Hardware-Oriented Security
  and Trust (HOST)}, 2013.

\bibitem{a2}
Kaiyuan Yang, Matthew Hicks, Qing Dong, Todd Austin, and Dennis Sylvester.
\newblock A2: Analog malicious hardware.
\newblock In {\em IEEE Symposium on Security and Privacy (S\&P)}, 2016.

\bibitem{zhang2017identifying}
Rui Zhang, Natalie Stanley, Christopher Griggs, Andrew Chi, and Cynthia
  Sturton.
\newblock Identifying security critical properties for the dynamic verification
  of a processor.
\newblock In {\em International Conference on Architectural Support for
  Programming Languages and Operating Systems}, ASPLOS, 2017.

\bibitem{zhang2020transys}
Rui Zhang and Cynthia Sturton.
\newblock Transys: Leveraging common security properties across hardware
  designs.
\newblock In {\em \textit{IEEE Symposium on Security and Privacy (S\&P)}},
  2020.

\bibitem{zhang2011ron}
Xuehui Zhang and Mohammad Tehranipoor.
\newblock Ron: An on-chip ring oscillator network for hardware trojan
  detection.
\newblock In {\em 2011 Design, Automation \& Test in Europe}, pages 1--6. IEEE,
  2011.

\bibitem{zhou2015detecting}
Boyou Zhou, Ronen Adato, Mahmoud Zangeneh, Tianyu Yang, Aydan Uyar, Bennett
  Goldberg, Selim Unlu, and Ajay Joshi.
\newblock Detecting hardware trojans using backside optical imaging of embedded
  watermarks.
\newblock In {\em Proceedings of IEEE Design Automation Conference (DAC)},
  2015.

\end{thebibliography}

\end{document}